\documentclass{aa}
\usepackage{epsfig,color,natbib}
\usepackage[dvips]{rotating}
\usepackage{url}
\usepackage{epsfig}
\usepackage{latexsym}
\newcount\longrefs
\def\aap{\ifnum\longrefs=1 {Astron.\ Astrophys.}\else 
                           {A\hbox{\rm \&}A}\fi}
\def\aapr{\ifnum\longrefs=1 {Astron.\ Astrophys.\ Rev.}\else 
                            {A\hbox{\rm \&}AR}\fi}
\def\aaps{\ifnum\longrefs=1 {Astron.\ Astrophys.\ Suppl.}\else 
                            {A\hbox{\rm \&}AS}\fi}
\def\aj{\ifnum\longrefs=1 {Astron.\ J.}\else 
                          {AJ}\fi} 
\def\ao{\ifnum\longrefs=1 {Applied Optics}\else 
                           {Appl.\ Opt.}\fi} 
\def\aspcs{\ifnum\longrefs=1 {Astron.\ Soc.\ Pacific Conf. Series}\else 
                           {ASP Conf.\ Ser.}\fi} 
\def\apj{\ifnum\longrefs=1 {Astrophys.\ J.}\else 
                           {ApJ}\fi} 
\def\apjl{\ifnum\longrefs=1 {Astrophys.\ J. Lett.}\else 
                            {ApJ}\fi} 
\def\aplett{\ifnum\longrefs=1 {Astrophys.\ J. Lett.}\else 
                            {ApJ}\fi} 
\def\apjs{\ifnum\longrefs=1 {Astrophys.\ J. Suppl.}\else 
                            {ApJS}\fi}
\def\apss{\ifnum\longrefs=1 {Astrophys.\ and Space Science}\else 
                            {Ap\hbox{\rm \&}SS}\fi}
\def\araa{\ifnum\longrefs=1 {Ann.\ Rev.\ Astron.\ Astrophys.}\else 
                            {ARA\hbox{\rm \&}A}\fi}
\def\azh{\ifnum\longrefs=1 {Astronomicheskii Zhurnal}\else 
                            {Astron.\ Zhur.}\fi}
\def\baas{\ifnum\longrefs=1 {Bull.\ Am.\ Astron.\ Soc.}\else 
                            {BAAS}\fi}
\def\bain{\ifnum\longrefs=1 {Bull.\ Astronom.\ Institutes Netherlands}\else
                            {Bull.\ Astr.\ Inst.\ Neth.}\fi}
\def\gca{\ifnum\longrefs=1 {Geochim.\ Cosmochim.\ Acta}\else 
                           {Geochim.\ Cosmochim.\ Acta}\fi}
\def\grl{\ifnum\longrefs=1 {Geophys.\ Res.\ Lett.}\else 
                           {Geoph.\ Res.\ Lett.}\fi}
\def\iaucirc{\ifnum\longrefs=1 {IAU Circulars}\else 
                          {IAU Circ.}\fi}
\def\ip{\ifnum\longrefs=1 {in press}\else 
                          {in press}\fi}
\def\jgr{\ifnum\longrefs=1 {J.\ Geophys.\ Res.}\else 
                           {J.\ Geophys.\ Res.}\fi}  
\def\jrasc{\ifnum\longrefs=1 {J.\ Royal Astron.\ Soc.\ Canada}\else 
                           {JRAS Can.}\fi}  
\def\mnras{\ifnum\longrefs=1 {Mon.\ Not.\ Roy.\ Astron.\ Soc.}\else 
                             {MNRAS}\fi} 
\def\memras{\ifnum\longrefs=1 {Roy. Astron. Soc., Mem.}\else 
                             {MmRAS}\fi} 
\def\nat{\ifnum\longrefs=1 {Nature}\else 
                           {Nat}\fi}
\def\pasj{\ifnum\longrefs=1 {Pub.\ Astron.\ Soc.\ Japan}\else 
                            {PASJ}\fi} 
\def\pasp{\ifnum\longrefs=1 {Pub.\ Astron.\ Soc.\ Pacific}\else 
                            {PASP}\fi} 
\def\physscr{\ifnum\longrefs=1 {Physica Scripta}\else 
                            {Phys.\ Scrip.}\fi} 
\def\planss{\ifnum\longrefs=1 {Planetary \& Space Science}\else 
                            {Plan. \& Space Sci.}\fi} 
\def\procspie{\ifnum\longrefs=1 {Proc.\ SPIE}\else 
                            {Proc.\ SPIE}\fi} 
\def\qjras{\ifnum\longrefs=1 {Quarterly J.\ Royal Astron.\ Soc.}\else 
                            {QJRAS}\fi} 
\def\sa{\ifnum\longrefs=1 {Soviet Astron..}\else 
                               {Sov.\ Astron.}\fi}
\def\skytel{\ifnum\longrefs=1 {Sky \& Telescope}\else 
                            {Sky \& Tel.}\fi} 
\def\solphys{\ifnum\longrefs=1 {Solar Phys.}\else 
                               {Solar Phys.}\fi}
\def\ssr{\ifnum\longrefs=1 {Space Science Rev.}\else 
                               {Space\ Sci.\ Rev.}\fi}

\def\dutch{\def\refname{Referenties}\def\abstractname{Samenvatting}%
  \def\bibname{Bibliografie}\def\chaptername{Hoofdstuk}%
  \def\appendixname{Bijlage}\def\contentsname{Inhoudsopgave}%
  \def\listfigurename{Lijst van figuren}\def\listtablename{Lijst van tabellen}%
  \def\indexname{Index}\def\figurename{Figuur}\def\tablename{Tabel}%
  \def\partname{Deel}\def\enclname{Bijlage(n)}\def\ccname{Ter attentie van}%
  \def\headtoname{Aan}\def\headpagename{Pagina}%
  \def\today{\number\day\space\ifcase\month\or januari\or februari\or maart\or%
     april\or mei\or juni\or juli\or augustus\or september\or oktober\or%
     november\or december\fi \space\number\year}%
  \typeout{
              >>>>> use hlatex209 for Dutch hyphenation <<<<< 
         }}
\newcounter{onefig} \newcounter{fignumber}
\newcount\nocaptions \newcount\nofigures \newcount\figwidth
\newcount\viewgraphs
  \def\paper{}  \def\figlabel{} 
\long\def\nextfig#1{\setcounter{figure}{\value{fignumber}}
  \addtocounter{fignumber}{1}
  \ifnum \viewgraphs=1 \newpage \pagestyle{empty} \fi 
  \ifnum\value{onefig}=0 #1 \fi                 
  \ifnum\value{onefig}=\value{fignumber} #1 \fi}
\def\figwidths#1#2{\ifnum \nocaptions=1 #2mm \else #1mm \fi}  
\def\paper#1{}  %% redefine for separate-figure identification line
\long\def\plotfig#1#2{\ifnum \nofigures=1 \else #2 \fi}
\long\def\captiontext#1{\ifnum \nofigures=1 \raggedright \fi 
   \ifnum \nocaptions=1 \paper
     \ifnum \viewgraphs=0 
       \newline  \mbox{}\hrulefill\mbox{} \newline 
       \newline label:~\{\figlabel\} 
     \fi 
     \else \ifnum \nofigures=0 \fi 
   #1 \fi}

\newcount\panelwidth \newcount\panelheight 
\newcount\bxmin \newcount\bymin \newcount\bxmax \newcount\bymax
\newcount\tbxmin \newcount\tbymin
\newcount\tpanelwidth \newcount\tpanelheight \newcount\tpdif
\def\panelsize #1,#2;{\panelwidth=#1 \panelheight=#2}  
\def\setbb #1,#2;#3,#4;#5,#6;{% UNITS: bp (from ghostview)
  \tbxmin=#1 \tbymin=#2    %% full box (axis titles) lower left corner
  \bxmin=#3 \bymin=#4      %% bare box (ticks only) lower left corner
  \bxmax=#5 \bymax=#6}     %% upper right corner
\def\barepanel #1{%
  \ifnum\panelheight=0 
    \tpdif=\bymax \advance\tpdif by -\bymin
    \multiply \tpdif by \panelwidth
    \tpanelheight=\tpdif
    \tpdif=\bxmax \advance\tpdif by -\bxmin
    \divide \tpanelheight by \tpdif
  \else \tpanelheight=\panelheight \fi
  \epsfig{file=#1,%
     bbllx=\bxmin bp,bblly=\bymin bp,bburx=\bxmax bp,bbury=\bymax bp,clip=,%
     width=\panelwidth mm,height=\tpanelheight mm}}
\def\labelypanel #1{% TeX permits only integer arithmetic, so bp and mm
  \ifnum\panelheight=0 
    \tpdif=\bymax \advance\tpdif by -\bymin
    \multiply \tpdif by \panelwidth
    \tpanelheight=\tpdif
    \tpdif=\bxmax \advance\tpdif by -\bxmin
    \divide \tpanelheight by \tpdif
  \else \tpanelheight=\panelheight \fi
  \tpdif=\bxmax \advance\tpdif by -\tbxmin
  \tpanelwidth=\panelwidth \multiply \tpanelwidth by \tpdif
  \tpdif=\bxmax \advance\tpdif by -\bxmin
  \divide \tpanelwidth by \tpdif
  \epsfig{file=#1,%
    bbllx=\tbxmin bp,bblly=\bymin bp,bburx=\bxmax bp,bbury=\bymax bp,%
    clip=,width=\tpanelwidth mm,height=\tpanelheight mm}}
\def\labelxpanel #1{%
  \ifnum\panelheight=0 
    \tpdif=\bymax \advance\tpdif by -\bymin
    \multiply \tpdif by \panelwidth
    \tpanelheight=\tpdif
    \tpdif=\bxmax \advance\tpdif by -\bxmin
    \divide \tpanelheight by \tpdif
  \else \tpanelheight=\panelheight \fi
  \tpdif=\bymax \advance\tpdif by -\tbymin
  \multiply \tpanelheight by \tpdif
  \tpdif=\bymax \advance\tpdif by -\bymin
  \divide \tpanelheight by \tpdif
  \epsfig{file=#1,%
    bbllx=\bxmin bp,bblly=\tbymin bp,bburx=\bxmax bp,bbury=\bymax bp,%
    clip=,width=\panelwidth mm,height=\tpanelheight mm}}
\def\labelxypanel #1{%
  \ifnum\panelheight=0 
    \tpdif=\bymax \advance\tpdif by -\bymin
    \multiply \tpdif by \panelwidth
    \tpanelheight=\tpdif
    \tpdif=\bxmax \advance\tpdif by -\bxmin
    \divide \tpanelheight by \tpdif
  \else \tpanelheight=\panelheight \fi
  \tpdif=\bxmax \advance\tpdif by -\tbxmin
  \tpanelwidth=\panelwidth \multiply \tpanelwidth by \tpdif
  \tpdif=\bxmax \advance\tpdif by -\bxmin
  \divide \tpanelwidth by \tpdif 
  \tpdif=\bymax \advance\tpdif by -\tbymin 
  \multiply \tpanelheight by \tpdif
  \tpdif=\bymax \advance\tpdif by -\bymin
  \divide \tpanelheight by \tpdif
  \epsfig{file=#1,%
    bbllx=\tbxmin bp,bblly=\tbymin bp,bburx=\bxmax bp,bbury=\bymax bp,%
    clip=,width=\tpanelwidth mm,height=\tpanelheight mm}}

\def\CC{\par \vspace*{-2ex} \footnotesize \baselineskip=8pt \begin{verbatim}}

\long\def\startignore #1\stopignore{}   %% use \startignore....\stopignore

\def\setlistparams{         
  \topsep=0.7ex                 %% ADAPT: parskip=0: 0.7;  parskip=1: -1.2ex
  \itemsep=0.7ex                %% space between items
  \leftmargini=3ex}             %% dashes at beginning of line 
\setlistparams                  %% recall after type changes 

\newcounter{alistindex}       %% problems: a)  b) etc

\newcounter{romenumnr}

\newlength{\minipagewidth}

\newsavebox{\boxcontent}
\newcommand{\ovalhead}[1]{
  \unitlength=1cm
  \sbox{\boxcontent}{\mbox{~~{#1}~~}}
  \begin{center}
    \ifdim\wd\boxcontent>6ex 
    \ifdim\wd\boxcontent<8cm 
    \begin{picture}(8,3) \thicklines     
      \put(4.0,0.8){\oval(8,1.6)} 
      \put(0.0,0.7){\parbox{8cm}{
         \begin{center} \usebox{\boxcontent} \end{center}}}
    \end{picture}
    \else \ifdim\wd\boxcontent<12cm 
    \begin{picture}(12,3) \thicklines     
        \put(6.0,0.8){\oval(12,1.6)} 
        \put(0.0,0.7){\parbox{12cm}{
           \begin{center} \usebox{\boxcontent} \end{center}}}
    \end{picture}
    \else
    \begin{picture}(16,3) \thicklines     
        \put(8.0,0.8){\oval(16,1.6)} 
        \put(0.0,0.7){\parbox{16cm}{
           \begin{center} \usebox{\boxcontent} \end{center}}}
    \end{picture}
    \fi \fi \fi
  \end{center}} 

\newcounter{headnr}            
\newcounter{subheadnr}[headnr]
\newcounter{subsubheadnr}[subheadnr]
\def\head #1\par{
  \stepcounter{headnr}                          %% sets subheadnr = 0 too 
  \vspace{2ex} \noindent                        %% 2ex = space above, no *
  {\bf \theheadnr~~~~#1}\\[1ex] \noindent}      %% 1ex = space below
\def\subhead #1\par{  
  \stepcounter{subheadnr}
  \vspace{1.3ex} \noindent
  {\bf \theheadnr.\arabic{subheadnr}~~~#1}\\[0.3ex] \noindent}
\def\subsubhead #1\par{
  \stepcounter{subsubheadnr}
  \vspace{1.0ex} \noindent
  {\bf \theheadnr.\arabic{subheadnr}.\arabic{subsubheadnr}~~~#1}\\ \noindent}

\font\dropfont= cmr12 scaled \magstep5
\def\dropcap#1#2{{\noindent
    \setbox0\hbox{\dropfont #1}\setbox1\hbox{#2}\setbox2\hbox{(}%
    \count0=\ht0\advance\count0 by\dp0\count1\baselineskip
    \advance\count0 by-\ht1\advance\count0by\ht2
    \dimen1=.5ex\advance\count0by\dimen1\divide\count0 by\count1
    \advance\count0 by1\dimen0\wd0
    \advance\dimen0 by.25em\dimen1=\ht0\advance\dimen1 by-\ht1
    \global\hangindent\dimen0\global\hangafter-\count0
    \hskip-\dimen0\setbox0\hbox to\dimen0{\raise-\dimen1\box0\hss}%
    \dp0=0in\ht0=0in\box0}#2}

\def\level #1 #2#3#4{$#1 \: ^{#2} \mbox{#3} ^{#4}$}   

\def\degree{\hbox{$^\circ$}}

\def\arcsec{\hbox{$^{\prime\prime}$}}

\def\kms{\hbox{km$\;$s$^{-1}$}}

\def\mathstacksym#1#2#3#4#5{\def#1{\mathrel{\hbox to 0pt{\lower 
    #5\hbox{#3}\hss} \raise #4\hbox{#2}}}}

\mathstacksym\lta{$<$}{$\sim$}{1.5pt}{3.5pt} % less than approximately
\mathstacksym\gta{$>$}{$\sim$}{1.5pt}{3.5pt} % greater than approximately
\mathstacksym\lrarrow{$\leftarrow$}{$\rightarrow$}{2pt}{1pt} % equilibrium
\mathstacksym\lessgreat{$>$}{$<$}{3pt}{3pt} %% less or greater

\bibpunct{(}{)}{;}{a}{}{,}
\begin{document}

\newcommand{\chitwo}{$rms$}
\newcommand{\chitwomin}{$rms_{\rm min}$}
\newcommand{\bvi}{$B, V, I_c$}
\newcommand{\gdor}{$\gamma$\,Dor}
\newcommand{\dsct}{$\delta$\,Sct}
\newcommand{\hipparcos}{{\sc hipparcos}}
\newcommand{\coralie}{{\sc coralie}}
\newcommand{\geneva}{{\sc geneva}}
\newcommand{\aurelie}{{\sc aur\'elie}}
\newcommand{\Mbol}{M$_{\mathrm{bol}}$}
\newcommand{\mbol}{m$_{\mathrm{bol}}$}
\newcommand{\ang}{\AA}
\newcommand{\Lsun}{L$_{\odot}$}
\newcommand{\Msun}{M$_{\odot}$}
\newcommand{\Ma}{$M_1$}
\newcommand{\Mb}{$M_2$}
\newcommand{\Ra}{$R_1$}
\newcommand{\Rb}{$R_2$}
\newcommand{\La}{$L_1$}
\newcommand{\Lb}{$L_2$}
\newcommand{\fap}{FAP-level}
\newcommand{\sn}{3.7\,S/N-level}
\newcommand{\Rsun}{R$_{\odot}$}
\newcommand{\cd}{$\rm{c\,d^{-1}}$}
\newcommand{\cy}{$\rm{c\,y^{-1}}$}
\newcommand{\teff}{T$_{\mathrm{eff}}$}
\newcommand{\va}{${<v>}$}
\newcommand{\vgamma}{$v_{\gamma}$}
\newcommand{\vai}{${<v>_{41ii}}$}
\newcommand{\vaa}{${<v>_{412.8}}$}
\newcommand{\vab}{${<v>_{413.0}}$}
\newcommand{\vahe}{${<v>_{412.1}}$}
\newcommand{\vsini}{$v \sin i$}
\newcommand{\vsinifou}{$v \sin i\,_{\rm Fourier}$}
\newcommand{\vsinifit}{$v \sin i\,_{\rm Fit}$}
\newcommand{\vb}{${<v^2>}$}
\newcommand{\vc}{${<v^3>}$}
\newcommand{\vd}{${<v^4>}$}
\newcommand{\ve}{${<v^5>}$}
\newcommand{\vf}{${<v^6>}$}
\newcommand{\siII}{Si\,II}
\newcommand{\logg}{$\log(g)$}
\newcommand{\HD}{HD\,}
\newcommand{\hd}{HD\,}
\newcommand{\HR}{HR\,}
\newcommand{\HIP}{HIP\,}
\newcommand{\hp}{H$_{\rm p}$}
\newcommand{\nua}{$\nu _1$}
\newcommand{\nub}{$\nu _2$}
\newcommand{\nuc}{$\nu _3$}
\newcommand{\nud}{$\nu _4$}
\newcommand{\nue}{$\nu _5$}
\newcommand{\nuf}{$\nu _6$}
\newcommand{\nug}{$\nu _7$}
\newcommand{\nuh}{$\nu _8$}
\newcommand{\nui}{$\nu _9$}
\newcommand{\nuj}{$\nu _{10}$}
\newcommand{\nuap}{$\nu _{1}^{\rm p}$}
\newcommand{\nubp}{$\nu _{2}^{\rm p}$}
\newcommand{\nucp}{$\nu _{3}^{\rm p}$}
\newcommand{\nudp}{$\nu _{4}^{\rm p}$}
\newcommand{\nuep}{$\nu _{5}^{\rm p}$}
\newcommand{\nufp}{$\nu _{6}^{\rm p}$}
\newcommand{\nugp}{$\nu _{7}^{\rm p}$}
\newcommand{\nuhp}{$\nu _{8}^{\rm p}$}
\newcommand{\nuip}{$\nu _{9}^{\rm p}$}
\newcommand{\nujp}{$\nu _{10}^{\rm p}$}
\newcommand{\nuas}{$\nu _{1}^{\rm s}$}
\newcommand{\nubs}{$\nu _{2}^{\rm s}$}
\newcommand{\nucs}{$\nu _{3}^{\rm s}$}
\newcommand{\nuds}{$\nu _{4}^{\rm s}$}
\newcommand{\nues}{$\nu _{5}^{\rm s}$}
\newcommand{\nufs}{$\nu _{6}^{\rm s}$}
\newcommand{\nugs}{$\nu _{7}^{\rm s}$}
\newcommand{\nuhs}{$\nu _{8}^{\rm s}$}
\newcommand{\nuis}{$\nu _{9}^{\rm s}$}
\newcommand{\nujs}{$\nu _{10}^{\rm s}$}
\newcommand{\pap}{$P _{1}^{\rm p}$}
\newcommand{\pbp}{$P _{2}^{\rm p}$}
\newcommand{\pcp}{$P _{3}^{\rm p}$}
\newcommand{\pdp}{$P _{4}^{\rm p}$}
\newcommand{\pep}{$P _{5}^{\rm p}$}
\newcommand{\pfp}{$P _{6}^{\rm p}$}
\newcommand{\pgp}{$P _{7}^{\rm p}$}
\newcommand{\php}{$P _{8}^{\rm p}$}
\newcommand{\pip}{$P _{9}^{\rm p}$}
\newcommand{\pjp}{$P _{10}^{\rm p}$}
\newcommand{\pas}{$P _{1}^{\rm s}$}
\newcommand{\pbs}{$P _{2}^{\rm s}$}
\newcommand{\pcs}{$P _{3}^{\rm s}$}
\newcommand{\pds}{$P _{4}^{\rm s}$}
\newcommand{\pes}{$P _{5}^{\rm s}$}
\newcommand{\pfs}{$P _{6}^{\rm s}$}
\newcommand{\pgs}{$P _{7}^{\rm s}$}
\newcommand{\phs}{$P _{8}^{\rm s}$}
\newcommand{\pis}{$P _{9}^{\rm s}$}
\newcommand{\pjs}{$P _{10}^{\rm s}$}
\newcommand{\triala}{$P _{\rm ellips}$}
\newcommand{\phipp}{$P _{\rm hipp}$}
\newcommand{\porb}{$P _{\rm orb}$}
\newcommand{\pa}{$P _1$}
\newcommand{\pb}{$P _2$}
\newcommand{\pc}{$P _3$}
\newcommand{\pd}{$P _4$}
\newcommand{\pe}{$P _5$}
\newcommand{\pf}{$P _6$}
\newcommand{\pg}{$P _7$}
\newcommand{\ph}{$P _8$}
\newcommand{\ka}{$K _1$}
\newcommand{\kb}{$K _2$}
\newcommand{\kc}{$K _3$}
\newcommand{\kd}{$K _4$}
\newcommand{\ke}{$K _5$}
\newcommand{\kf}{$K _6$}
\newcommand{\kg}{$K _7$}
\newcommand{\kh}{$K _8$}
\newcommand{\nuorb}{$\nu _{\rm orb}$}
\newcommand{\sig}{$\sigma$}
\newcommand{\sign}{$\sigma _N$}
\newcommand{\siga}{$\sigma _1$}
\newcommand{\sigb}{$\sigma _2$}
\newcommand{\sigc}{$\sigma _3$}
\newcommand{\sigd}{$\sigma _4$}
\newcommand{\sige}{$\sigma _5$}
\newcommand{\siglim}{$\sigma _{\rm lim}$}
\newcommand{\sigres}{$\sigma _{\rm res}$}
\newcommand{\siglam}{$\sigma _{\lambda}$}
\newcommand{\po}{$p_o$\,}
\newcommand{\gamlm}{$\gamma(\ell_1,m_1)$}
\newcommand{\siglm}{$\Gamma_\ell^m$}
\newcommand{\el}{$\eta_\ell$}
\newcommand{\vp}{$v_p$}
\newcommand{\vpK}{$v_p K$}
\newcommand{\vomega}{$v_{\Omega}$}
\newcommand{\veq}{$v_{\rm eq}$}
\newcommand{\nurot}{$\nu_{\Omega}$}
\newcommand{\vrad}{$v_{\rm rad}$}
\newcommand{\vcrit}{$v_{\rm crit}$}
\newcommand{\vrmax}{$v_{\rm r,max}$}
\newcommand{\vtmax}{$v_{\rm t,max}$}
\newcommand{\nucorot}{$\nu_{\rm co}$}
\newcommand{\nuspin}{$\eta$}
\newcommand{\fout}{$s_{f}$}
\newcommand{\hl}{$H_l$}
\newcommand{\Ba}{$B_1$}
\newcommand{\Bb}{$B_2$}
\newcommand{\Va}{$V_1$}
\newcommand{\vup}{$v_{\rm tot}$}

\title{A spectroscopic study of southern (candidate) $\gamma$ Doradus stars\thanks{Based on observations collected with the {\sc coralie} spectrograph attached to the Euler Telescope of the Geneva Observatory situated at La Silla (Chile)}} 
\subtitle{I. Time series analysis}

\author{P.~De Cat$^{1,2}$ \and L.~Eyer$^3$ \and J.~Cuypers$^{1}$ \and
  C.~Aerts$^{2,4}$ \and B.~Vandenbussche$^2$ \and
  K.~Uytterhoeven$^{2,5}$ \and K.~Reyniers$^2$ \and K.~Kolenberg$^{2,6}$ \and
  M.~Groenewegen$^2$ \and G.~Raskin$^{2,7}$ \and T.~Maas$^{2}$ \and S.~Jankov$^{8,9}$}

\offprints{P. De Cat}
\mail{peter@oma.be}

\institute{$^1$Royal Observatory of Belgium, Ringlaan 3, B-1180 Brussel, Belgium\\
$^2$Instituut voor Sterrenkunde, Katholieke Universiteit Leuven, Celestijnenlaan 200 B, B-3001 Leuven, Belgium\\
$^3$Observatoire de Gen\`eve, CH-1290 Sauverny, Switzerland\\
$^4$Department of Astrophysics, Radboud University Nijmegen, PO Box 9010, 6500 GL Nijmegen, the Netherlands\\
$^5$Centre for Astrophysics, University of Central Lancashire, Preston, PR1 2HE, United Kingdom\\
$^6$Institut f\"ur Astronomie, Universit\"at Wien, T\"urkenschanzstrasse 17, 1180 Wien, Austria\\
$^7$Mercator Telescope, Calle Alvarez de Abreu 70, E-38700 Santa Cruz de La
Palma, Spain\\
$^8$Observatoire de la C\^ote d'Azur, D\'epartement Gemini, UMR 6203, 06304 Nice Cedex 4, France\\
$^9$Laboratoire Univ. d'Astroph. de Nice (LUAN), UMR 6525, Parc Valrose, 06108 Nice Cedex 02, France}

\date{Received 18 June 2005 / Accepted 27 October 2005}

%%%%%%%%%%%%%%%%
\abstract{ We present results of a spectroscopic study of 37 southern
(candidate) \gdor adus stars based on \'echelle spectra.
The observed spectra were cross-correlated with the standard template
spectrum of an F0-type star for an easier detection of binary and intrinsic
variations. 
We identified 15 objects as spectroscopic binaries, including 7 new ones, and
another 3 objects are binary suspects.
At least 12 objects show composite spectra. 
We could determine the orbital parameters for 9 binaries, of which 4 turn out
to be ellipsoidal variables.
For 6 binaries, we estimated the expected time-base of the orbital variations.
Clear profile variations are observed for 17 objects, pointing towards
stellar pulsation.
For 8 of them, we have evidence that the main spectroscopic and photometric
periods coincide.
Our results, in combination with prior knowledge from the literature, lead to
the classification of 10 objects as new {\it bona-fide} \gdor adus stars, 1
object as new {\it bona-fide} $\delta$ Scuti star, and 8 objects as constant
stars. 
Finally, we determined the projected rotational velocity with two independent
methods. 
The resulting \vsini\ values range from 3 to 135 \kms.
For the bona-fide $\gamma$ Doradus stars, the majority has \vsini\ below
60~\kms. 

\keywords{Stars : variables : general -- Stars: oscillations -- Line: profiles} }
%%%%%%%%%%%%%%

\maketitle
\markboth{De Cat et al.: A spectroscopic study of southern (candidate) \gdor adus stars. I.}
         {De Cat et al.: A spectroscopic study of southern (candidate) \gdor adus stars. I.}

\section{Introduction}
\label{intro}

The $\gamma$ Doradus stars (\gdor\ stars hereafter) are a group of variable
early F-type stars situated along the main-sequence. 
They were recognized as a separate group of pulsating stars one decade ago
\citep{Balona1994MNRAS.270..905B}.
Their observed variations with typical periods between 0.3 and 3
days are due to $g$-mode pulsations which carry asteroseismic
information about the central stellar regions. 
Their pulsations are driven by a flux blocking mechanism at the base of their
convective envelope (e.g. \citealt{Guzik2000STIN...0205964G,
Loffler2002ASPC..259..508L, Warner2003ApJ...593.1049W,
Dupret2004A&A...414L..17D}).
Hopes of progress in this field boosted many astronomers to observe \gdor\
stars to allow further constraints on both the observed pulsational
characteristics and the observed \gdor\ instability strip in the HR diagram. 

The photometric measurements of the satellite mission {\sc hipparcos} led to
an impressive growth in the number of \gdor\ candidates: from 11
\citep{Krisciunas1995IBVS.4195....1K} to 127 \citep{Handler2002}. 
The number of bona-fide members grew from 6
\citep{Krisciunas1995IBVS.4195....1K} over 42 \citep{Handler2002} to 54
\citep{Henry2005AJ....129.2815H}. 
The reason to divide the list into two classes, the bona-fide members and the
candidates (hereafter respectively ``bf \gdor'' and ``cand \gdor''), arises
from the fact that the variability of the candidate stars can eventually be
attributed to diverse phenomena: eclipsing binaries, ellipsoidal variations,
spots, or $\delta$\,Scuti (hereafter ``\dsct'') pulsations (due to aliasing problems). 
Therefore the confirmation that a star belongs to the \gdor\ stars group
requires often extensive observations, preferably including spectroscopic
ones. 
 
Since most of the bf \gdor\ stars are multi-periodic, the observed
light curves are generally rather complex.
Long-term monitoring is not only needed to reveal the frequency spectrum, but
also to check the stability of the observed periods/phases and amplitudes.
Indeed, an excess of scatter around maximum brightness has already been
observed in some bf \gdor\ stars 
\citep{Zerbi1997MNRAS.290..401Z, Zerbi1997MNRAS.292...43Z,
  Zerbi1999MNRAS.303..275Z, Henry2001AJ....122.3383H,
  Fekel2003AJ....125.2156F}.

In recent years, large scale photometric and spectroscopic follow-up
campaigns have already been undertaken by other researchers, mainly for
{\it northern} cand \gdor\ stars (e.g. \citealt{Kaye1999AJ....118.2997K,
Henry2001AJ....122.3383H, Henry2002PASP..114..988H,
Henry2003AJ....126.3058H,Fekel2003AJ....125.2196F,
Mathias2004A&A...417..189M,Henry2005AJ....129.2026H,Henry2005AJ....129.2815H}). 
In this series of papers, we report on the results obtained for a sample of 37
{\it southern} confirmed and cand \gdor\ stars with time-series of
new spectroscopic data gathered in 1998--2003 with the \coralie\ spectrograph
attached to the 1.2-m Euler telescope located in La Silla (Chile).
Paper\,I (this paper) is devoted to the time series analysis of the data. 
These results are used for the orbital and variability classification of our
targets.
In Paper\,II (Bruntt et al., in preparation), the chemical abundances are
determined for the slow rotators within our sample.
%\\
A complementary photometric study has been realised for a slightly different
sample of 37 {\it southern} (cand) \gdor\ stars based on new photometric
observations in the Johnson-Cousins \bvi\ filters obtained in 1999--2000 with
the {\sc modular} photometer attached to the 0.5-m {\sc saao} telescope
located in  Sutherland (South-Africa). 
The first results of this campaign were discussed by \citet{Bouckaert2001UT}
and \citet{Eyer2002oapb.conf..203E}, while Eyer et al. (in preparation) are
preparing the full report.
%\\
Furthermore, the 1.2-m Mercator telescope located on the Canary Island La
Palma (a twin of the Euler telescope) has become operational since 2001 and is
currently used to monitor, amongst others, {\it northern} (cand) \gdor\ stars
photometrically in the Geneva $U, B_1, B, B_2, V_1, V, G$ passbands.
The results obtained after 18 months observations are discussed by
\citet{DeRidder2004ASPC..310..263D}. 
The results after 3 years of observation will become available soon (Cuypers
et al., in preparation).

This paper is organised as follows.
The data-sets obtained with the \coralie\ spectrograph, which are
described in Section\,\ref{data}, are used for three main purposes:
(1) to determine which of the objects are binaries, and, if possible, to
determine their orbits (Section\,\ref{binaries}), 
(2) to confirm/reject the classification of bf or cand \gdor\
stars by studying the intrinsic variability of the spectra
(Section\,\ref{intrinsic}), 
(3) to determine accurate values of the projected rotational velocity \vsini\
for each target star (Section\,\ref{methods}).
In Section\,\ref{discussion}, we end with our general conclusion and future
prospects.

\section{Description of the data-sets}
\label{data}

\begin{table*}
\caption{\label{targets} 
Overview of the 37 targets discussed in this paper. 
For each object, we give: the HD number (1), the \hipparcos\ number (2), the
Johnson $V$ magnitude (3) and the spectral classification (4) as found
in the {\sc simbad} astronomical database; the number of \coralie\ spectra (5)
and their total time span expressed in days (6); 
the main \hipparcos\ period in days (7); the orbital classification
(8); an indication for the observation of cross-correlation profile variations
(9); the variability classification (10); and the mean \vsini\ values obtained
by least-squares fitting with synthetic profiles for the primary (11) and
secondary (12) components. 
The errors are given between brackets in units of the last decimal.
The targets which are in common with those of the sample of Eyer et al. (in
preparation) are indicated with $\ast$ in column (1). 
The stars which were already classified as confirmed or suspected binaries are
indicated with $^+$ in column (8).
The stars which were already classified as bf \gdor\ stars are
indicated with $\degree$ in column (10). 
}
\begin{center}
\tabcolsep=4.5pt
\begin{tabular}{llclrrlllllcc} \hline 
&&&&&&&& \\ 
[-8pt]
HD        & HIP     & $m_V$ & $SpT$    & $\#$ & \multicolumn{1}{c}{$\Delta T$} & \multicolumn{1}{c}{\phipp} && binarity & CPVs & variability  &\multicolumn{2}{c}{$v\sin i$} \\ %\multicolumn{1}{c}{$\Delta\nu$} &
&& \multicolumn{1}{c}{\tiny(mag)} &&& \multicolumn{1}{c}{\tiny(days)} & \multicolumn{1}{c}{\tiny(days)} &&&&& \multicolumn{2}{c}{\tiny(\kms)}  \\ %\multicolumn{1}{c}{\tiny(\cd)} &      
\multicolumn{1}{c}{(1)}&\multicolumn{1}{c}{(2)}&(3)&\multicolumn{1}{l}{(4)}&(5)&\multicolumn{1}{c}{(6)}&\multicolumn{1}{c}{(7)}&&(8)&(9)&(10) & \multicolumn{1}{c}{(11)} & \multicolumn{1}{c}{(12)} \\ [2pt] \hline \\ [-8pt]%\multicolumn{1}{c}{(7)}&
  5590      &   4481  &  9.21 & F2\,V    &  16  &  1355      &          &            & SB2    &     & constant                  & 103(3)   & 3.4(0.3) \\%  0.0003 &
  7455      &   5745  &  8.47 & G3\,V    &  19  &  1723      &          &            &        & no  & constant                  &   3.3(3) &          \\%  0.0002 &
  8393      &   6387  &  9.49 & F7w      &  14  &  1136      &          &            & SB2    &     & constant                  &   4.1(3) &   3.7(2) \\%  0.0003 &
 10167$\ast$&   7649  &  6.67 & F0\,V    &  15  &  1721      &          &            & SB2$^+$&     & cand \gdor                &   4.5(5) &   4.9(6) \\%  0.0002 & 
 12901$\ast$&   9807  &  6.74 & F0       &  50  &  1721      & 2.18:    &{\tiny [2]} &        & yes & bf \gdor$\degree$         &  64(3)   &          \\%  0.0004 &
 14940$\ast$&  11192  &  6.68 & F0\,IV/V &  63  &   685      & 0.50038  &{\tiny [1]} &        & yes & bf \gdor                  &  39(4)   &          \\%  0.02   & 
 22001      &  16245  &  4.71 & F5\,IV/V &  18  &  1719      &          &            &        & no  & constant                  &  12(2)   &          \\%  0.0003 &
 26298$\ast$&  19383  &  8.16 & F0/F2\,V &   9  &   332      &          &            & SB1?   &     & cand \gdor?               &  50.9(5) &          \\%  0.0010 &
 27290      &  19893  &  4.26 & F4\,III  &  19  &  1395      & 0.757031 &{\tiny [1]} &        & yes & bf \gdor$\degree$(/\dsct?)&  55(2)   &          \\%  0.0003 &
 27377$\ast$&  20036  &  7.4  & F0\,V    &   3  &    13      & 2.8482   &{\tiny [3]} & SB2    &     & cand \gdor?               &   5.5(4) &   5.2(6) \\%  0.03   & 
 27604      &  20109  &  6.08 & F7\,IV/V &   3  &     6      &          &            & VB?    &     & constant                  &  67.5(4) &   4.3(3) \\%  0.07   & 
 33262      &  23693  &  4.71 & F7\,V    &  19  &  1722      & 0.28603  &{\tiny [3]} &        & no  & constant                  &  10.1(10)&          \\%  0.0003 &
 34025$\ast$&  24215  &  7.87 & F2\,IV   &   8  &   302      & 0.62221  &{\tiny [1]} & SB2    & yes & bf \gdor                  &  62.8(11)&  39(2)   \\%  0.0012 & 
 35416$\ast$&  25183  &  7.53 & F3\,V    &  11  &   625      &          &            & SB2    &     & cand \gdor?               &  6.6(?)  &  6.0(?)  \\%  0.0014 & 
 40745$\ast$&  28434  &  6.21 & F2\,IV   &   8  &   334      & 0.82415  &{\tiny [1]} &        & yes & bf \gdor                  &  39(2)   &          \\%  0.0011 & 
 41448$\ast$&  28778  &  7.60 & A9\,V    &  12  &    10      & 0.419912 &{\tiny [1]} &        & yes & bf \gdor                  & 106(3)   &          \\%  0.05   & 
 48501$\ast$&  32144  &  6.26 & F2\,V    &  34  &  1493      &          &            &        & yes & bf \gdor$\degree$         &  44(4)   &          \\%  0.0003 &
 65526$\ast$&  39017  &  6.98 & A3       &   2  &     1      & 1.28798  &{\tiny [1]} &        & yes & bf \gdor$\degree$         &  56(4)   &          \\%  0.3    &
 81421$\ast$&  46223  &  7.01 & A3       &  63  &   384      & 0.489724 &{\tiny [1]} & SB2$^+$&     & constant                  &  62.0(4) &  42(4)   \\%  0.03   & 
 85964$\ast$&  48580  &  7.52 & F3\,IV/V &  11  &   381      & 0.62425  &{\tiny [1]} & SB1    &     & constant                  &  65(2)   &          \\%  0.0012 & 
110379      &         &  3.65 & F0\,V    &   8  &    15      &          &            &        &     & cand \gdor                &  23.9(6) &          \\%  0.03   & 
110606$\ast$&  62105  &  7.8  & F2\,V    &   7  &    75      & 0.977    &{\tiny [2]} & SB2$^+$&     & cand \gdor                &   5.8(4) &  10(2)   \\%  0.005  & 
111709$\ast$&  62774  &  9.27 & A3:m...  &   5  &    78      & 1.18567  &{\tiny [1]} & SB2    & yes & cand \gdor/CP star?       &  61(1)   &  25(4)   \\%  0.005  & 
111829$\ast$&         &  9.49 & A1\,IV/V &   2  &     0      &          &            & SB2?   &     & cand \gdor?               &  47(2)   &   ?      \\%         & 
112685$\ast$&  63372  &  7.85 & F0\,V    &  24  &   385      & 0.61645  &{\tiny [1]} &        & yes & bf \gdor                  &  70(3)   &          \\%  0.004  & 
112934$\ast$&  63491  &  6.6  & A9\,V    &   5  &    79      & 0.8:     &{\tiny [2]} &        &     & cand \gdor                &  70.5(7) &          \\%  0.005  & 
125081$\ast$&  69848  &  7.35 & F2\,II   &   7  &    77      & 0.153981 &{\tiny [1]} &        & yes & bf \dsct                  &   5.6(5) &          \\%  0.005  & 
126516$\ast$&  70566  &  8.2  & F3\,V    &   9  &    76      & 0.493    &{\tiny [2]} & SB1$^+$&     & cand \gdor                &   3.8(3) &          \\%  0.02   & 
135825$\ast$&  74825  &  7.31 & F0       &  17  &   450      & 0.76053  &{\tiny [1]} &        & yes & bf \gdor                  &  38(5)   &          \\%  0.0009 & 
147787$\ast$&  80645  &  5.28 & F4\,IV   &  10  &   377      & 1.45557  &{\tiny [1]} & SB2$^+$&     & cand \gdor                &  7.6(?)  & 25(?)    \\%  0.0010 & 
149989$\ast$&  81650  &  6.30 & A9\,V    &  10  &   379      & 0.42658  &{\tiny [1]} &        & yes & bf \gdor                  & 134(3)   &          \\%  0.0012 & 
167858$\ast$&  89601  &  6.62 & F2\,V    &   7  &     3      & 1.30700  &{\tiny [1]} & SB1$^+$&     & bf \gdor$\degree$         &   5.0(2) &          \\%  0.16   &
187028$\ast$&  97590  &  7.60 & F0\,V    &  21  &   179      & 0.69532  &{\tiny [1]} &        & yes & bf \gdor                  &  85(3)   &          \\%  0.03   & 
209295$\ast$& 108976  &  7.5  & A9/F0\,V &  61  &  1399      & 0.885274 &{\tiny [1]} & SB1$^+$& yes & bf \gdor/\dsct$\degree$   &  86(3)   &          \\%  0.0012 &
214291      & 111718  &  6.54 & F7\,V    &   5  &  1329      & 0.87125  &{\tiny [1]} & SB2$^+$&     & cand \gdor?               &  67.6(8) &  64.7(7) \\%  0.0005 & 
216910$\ast$& 113402  &  6.70 & F2\,IV   &  11  &  1718      & 0.69349  &{\tiny [1]} &        & yes & bf \gdor                  &  92(3)   &          \\%  0.0002 & 
218225$\ast$& 114127  &  8.72 & F3\,IV   &  14  &  1327      & 0.86679  &{\tiny [1]} &        & yes & bf \gdor                  &  59(3)   &          \\%  0.007  & 
[2pt] \hline
\end{tabular}\\
{\tiny [1] \citet{ESA1997}; [2] \citet{Handler1999MNRAS.309L..19H}; [3] \citet{Koen2002MNRAS.331...45K}}
\end{center}
\end{table*}

\begin{table}
\caption{\label{log} 
Logbook of the spectroscopic observations.
The column $\#$ denotes the number of \coralie\ spectra taken for the project of
southern (cand) \gdor\ stars during each of the 14 observation runs.
}
\begin{center}
\begin{tabular}{lcl} \hline 
& \\ [-8pt]
Observation run            &$\#$ & Observer               \\ [2pt] \hline \\ [-8pt]
26/11/1998 -- 06/12/1998   &  40 & B. Vandenbussche       \\
06/04/1999 -- 15/04/1999   &  42 & K. Kolenberg           \\
28/05/1999 -- 04/06/1999   &  69 & L. Eyer                \\
15/10/1999 -- 27/10/1999   & 108 & B. Vandenbussche       \\
23/11/1999 -- 06/12/1999   &  27 & B. Vandenbussche       \\
16/02/2000 -- 29/02/2000   &  50 & K. Uytterhoeven        \\
17/04/2000 -- 25/04/2000   &  50 & K. Uytterhoeven        \\
20/06/2000 -- 06/07/2000   &  55 & T. Reyniers            \\
03/08/2000 -- 16/08/2000   &  73 & B. Vandenbussche       \\
28/09/2000 -- 11/10/2000   &  74 & K. Uytterhoeven        \\
13/11/2001 -- 26/11/2001   &   3 & G. Raskin              \\
12/07/2002 -- 25/07/2002   &   8 & B. Vandenbussche       \\
21/10/2002 -- 03/11/2002   &   2 & T. Maas                \\
18/12/2002 -- 29/12/2002   &  24 & M. Groenewegen         \\ [2pt] \hline \\ [-8pt]
26/11/1998 -- 29/12/2002   & 625 &                        \\ [2pt] \hline
\end{tabular}
\end{center}
\end{table}

\begin{table*}
\caption{\label{elements} 
Overview of the orbital elements determined for the ellipsoidal variables
(\hd34025, \hd81421, \hd214291, \hd85964) and the binaries with a (cand)
\gdor\ component (\hd10167, \hd147787, \hd126516, \hd167858, \hd209295) in
our sample of 37 targets.
The errors are given between brackets in units of the last decimal.
The values that were fixed are given in {\it italics}.
The targets for which an orbit was already known in the literature are
indicated with $\ast$.
}
\begin{center}
\tabcolsep=2.5pt
\begin{tabular}{lccccccccc} \hline
& \\ [-8pt]
                                        & \hd10167     & \hd34025     & \hd81421     & \hd85964    & \hd126516    & \hd147787    & \hd167858    & \hd209295   & \hd214291  \\
&&{\tiny (ellipsoidal)}&{\tiny (ellipsoidal)} &{\tiny (ellipsoidal)} &&$\ast$&$\ast$&$\ast$&{\tiny (ellipsoidal)} \\ [2pt] \hline \\ [-8pt]
\porb\ {\tiny (days)}                   & 9.3199(2)    & {\it 1.24442}&{\it 0.97948} &{\it 1.24850}& 2.1245(5)    & 39.86(8)     & 4.48510(13)  & 3.10573(2)  & 1.74247(2) \\
$v_{\gamma}$ {\tiny (\kms)}             & 5.2(2)       & 28.1(8)      & 9.3(4)       & 4.1(4)      & -20.6(13)    & -4.9(6)      & -28.3(2)     & -20.5(2)    & -13.9(6)   \\
$e$                                     & {\it 0.0}    & {\it 0.0}    & {\it 0.0}    & {\it 0.0}   & {\it 0.0}    & {\it 0.59}   & {\it 0.0}    & 0.324(5)    & {\it 0.0}  \\
$t(\tau)$ {\tiny (HJD-2450000)}         &              &              &              &             &              & 1312(2)      &              & 2865.097(12)&            \\     
$\omega$ {\tiny ($^{\circ}$)}           &              &              &              &             &              & 97(10)       &              & 33(1)       &            \\ 
$K_1$ {\tiny (\kms)}                    & 39.7(4)      & 81(2)        & 87.8(7)      & 65.5(5)     & 33(1)        & 61(72)       & 6.3(2)       & 52.8(3)     & 111(14)    \\
$K_2$ {\tiny (\kms)}                    & 41.8(4)      & 135(3)       & 164.7(7)     &             &              & 67(79)       &              &             & 111(14)    \\
$a_1 \sin i$ {\tiny (A.U.)}             & 0.0340(3)    & 0.0093(2)    & 0.00790(6)   & 0.00752(6)  & 0.0065(3)    & 0.2(2)       & 0.00261(8)   & 0.01426(8)  & 0.018(2)   \\
$a_2 \sin i$ {\tiny (A.U.)}             & 0.0358(3)    & 0.0154(3)    & 0.01483(6)   &             &              & 0.2(3)       &              &             & 0.018(2)   \\
$M_1 \sin ^{3} i $ {\tiny ($M_{\odot}$)}& 0.268(11)    & 0.81(9)      & 1.07(3)      &             &              & 2(12)        &              &             & 1.0(5)     \\
$M_2 \sin ^{3} i $ {\tiny ($M_{\odot}$)}& 0.254(8)     & 0.49(4)      & 0.568(15)    &             &              & 2(9)         &              &             & 1.0(4)     \\
$f(M)$ {\tiny ($M_{\odot}$)}            &              &              &              & 0.0364(8)   & 0.0080(9)    &              & 0.00012(2)   & 0.0401(7)   &            \\ [2pt] \hline \\ [-8pt]
$rms$ {\tiny (\kms)}                    & 0.99         & 1.43         & 2.60         & 0.87        & 1.97         & 0.98         & 0.89         & 1.03        & 0.67       \\ [2pt] \hline
\end{tabular}
\end{center}
\end{table*}

\begin{figure*}
\begin{center}
\resizebox{7.42cm}{!}{\rotatebox{270}{\includegraphics{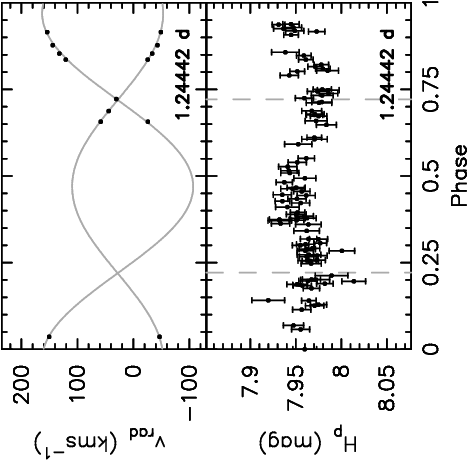}}}
\resizebox{5.2cm}{!}{\rotatebox{270}{\includegraphics{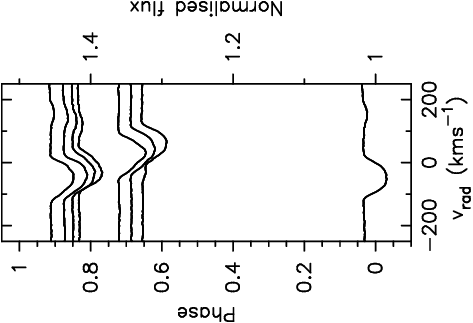}}}
\end{center}
\caption{\label{hd34025fase} 
Phase diagram of the radial velocity \vrad\ (top left) and the Hipparcos \hp\
measurements (bottom left) of \hd34025 with the period as given in the bottom right corner.
The reference epoch is HJD 2450000.
The dashed lines in the bottom left panel denote the phase at which \vrad\,=
$v_{\gamma}$.
In the right panel, a selection of observed cross-correlation profiles are
shown as a function of orbital phase.
}
\end{figure*}

\setcounter{figure}{2}

\setcounter{figure}{3}

\begin{figure*}
\begin{center}
\resizebox{7.42cm}{!}{\rotatebox{270}{\includegraphics{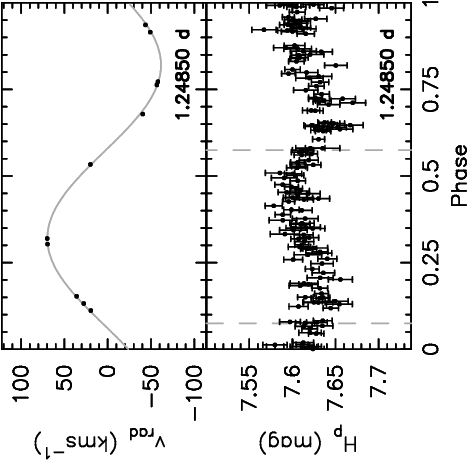}}}
\resizebox{5.20cm}{!}{\rotatebox{270}{\includegraphics{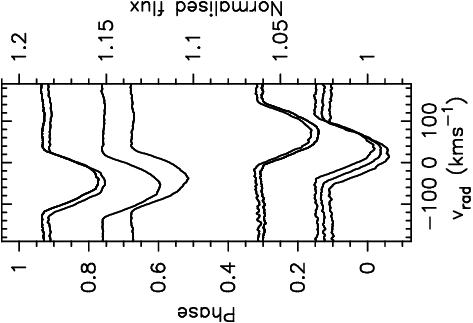}}}
\end{center}
\caption{\label{hd85964fase} 
Same as Fig.\,\ref{hd34025fase}, but for \hd85964.
}
\end{figure*}

\begin{figure*}
\begin{center}
\resizebox{7.42cm}{!}{\rotatebox{270}{\includegraphics{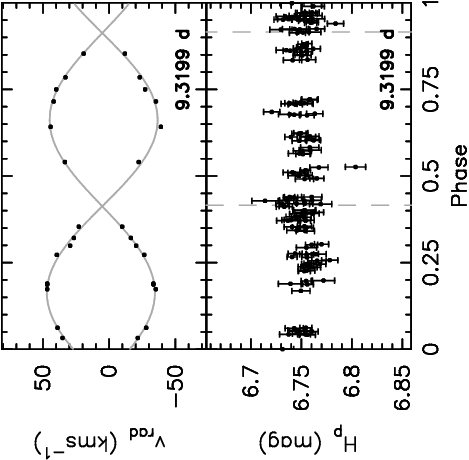}}}
\resizebox{5.20cm}{!}{\rotatebox{270}{\includegraphics{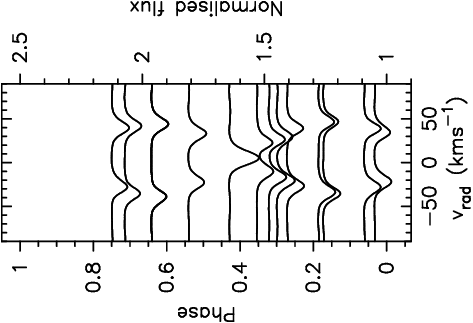}}}
\end{center}
\caption{\label{hd10167fase} 
Same as Fig.\,\ref{hd34025fase}, but for \hd10167.
}
\end{figure*}

\setcounter{figure}{6}

\begin{figure*}
\begin{center}
\resizebox{7.42cm}{!}{\rotatebox{270}{\includegraphics{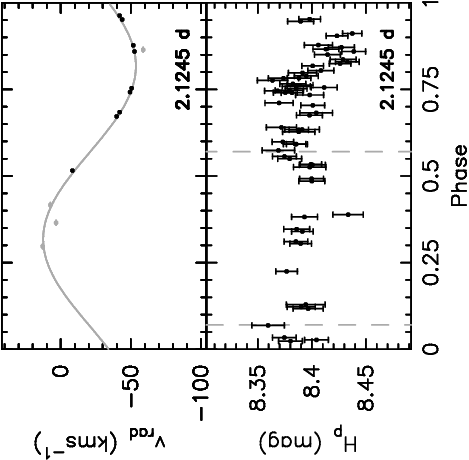}}}
\resizebox{5.20cm}{!}{\rotatebox{270}{\includegraphics{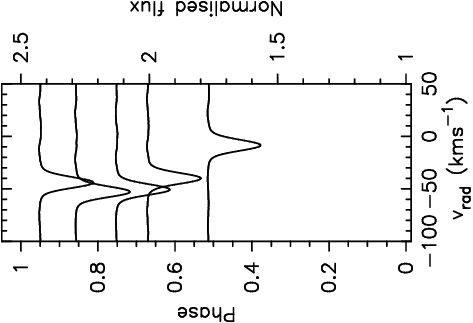}}}
\end{center}
\caption{\label{hd126516fase} 
Same as Fig.\,\ref{hd34025fase}, but for \hd126516.
}
\end{figure*}

\setcounter{figure}{8}

\setcounter{figure}{9}

A sample of 37 {\it southern} confirmed and cand \gdor\
stars were selected for spectroscopic monitoring from the lists given by
\citet{Eyer1998PhDT.........8E}, \citet{Aerts1998A&A...337..790A} and
\citet{Handler1999MNRAS.309L..19H}. 
For each object, the HD number, the {\sc hipparcos} number, the
Johnson $V$ magnitude and the spectral classification as found
in the {\sc simbad} astronomical database are given in columns (1)--(4) of
Table\,\ref{targets}. 
The 28 targets which are in common with those that were selected for
photometric monitoring (Eyer et al., in preparation) are indicated with $\ast$.

Between Nov.~1998 and Dec.~2002, we gathered high-resolution spectra during 14
observation campaigns (Table\,\ref{log}) with the \coralie\ spectrograph
attached to the 1.2-m {\sc euler} telescope located in La Silla (Chile).
{\sc coralie} is a 2-fiber-fed \'echelle spectrograph (2\arcsec\ fib res on the
object and sky, respectively), which covers the 388--681~nm region in 68
orders with a spectral resolution of 50~000.
During each night, several exposures with a tungsten lamp were taken to
measure the relative pixel sensitivity variation of the CCD.
Exposures with a thorium lamp were taken for the wavelength calibration.
The science exposure times were chosen for each object individually to result
in a S/N ratio of about 100.
The resulting number of science exposures for each target and their total time
span expressed in days are respectively given in columns (5) and (6) of
Table\,\ref{targets}. 

%%%%%%%%%%%
%\subsection{Reductions}
%%%%%%%%%%%
Right after the science exposure is taken, an optimised pipeline reduction is
carried out with the {\sc inter-tacos} software.
%
%%%%%%%%%%%
%\subsection{Cross-correlation}
%%%%%%%%%%%
%
The wavelength calibrated 2-D science spectra were used without either
rebinning or merging the orders to calculate normalized cross-correlation
functions (CCFs) in the solar system barycentric reference frame in the same
way as described by \citet{Baranne1996A&AS..119..373B}.  
For all objects, we took the standard template spectrum of an F0-type star as
correlation mask. 
Given their S/N ratio, time-series of CCFs allow a much easier detection of
(double-lined) binaries (see Section\,\ref{binaries}) and/or correlation
profile variations (CPVs; see Section\,\ref{intrinsic}) compared to
time-series of individual spectral lines.

%%%%%%%%%%%
%\subsection{Radial velocity and vsini}
%%%%%%%%%%%

The CCFs were used to determine the radial velocity (\vrad).
In a first approach, \vrad\ was calculated by fitting the CCF with a Gaussian
distribution function. 
This approach was used only for double-lined objects in phases where the
central parts of the lines of the different components are well enough
separated in velocity.  
In a second approach, the first normalized velocity moment (as defined by
\citealt{Aerts1992A&A...266..294A}) of the CCF is used to calculate \vrad.
This approach, which is favourable in case of pronounced CPVs, was used for all
our single-lined objects.
The accuracy of the obtained \vrad\ values remains well below 1~\kms\
(cf. Table\,\ref{sigma}).  
The \vrad\ time-series were used to search for orbital/intrinsic
periods in the observed variations. 
To check the consistency between the spectroscopic and photometric
variability, we additionally used the \hipparcos\ \hp\ measurements
\citep{ESA1997}.
The main \hipparcos\ period in days is given in column (7) of
Table\,\ref{targets}.  

Since the spectroscopic data of both \hd12901 and \hd48501 were already
analysed in detail by \citet{Aerts2004A&A...415.1079A}, these objects will not
be discussed any more in the following sections although they do belong to the
sample.  

\section{Orbital variations}
\label{binaries}

For the objects for which we have enough data, the time-series of
\vrad\ are subjected to a period search.
Two period search algorithms were used: 
(1) Phase Dispersion Minimization (PDM hereafter;
\citealt{Stellingwerf1978ApJ...224..953S}),
% which is well suited to detect
%non-sinusoidal variations which can be expected for e.g. eccentric binaries,
and (2) the Lomb-Scargle periodogram (Scargle hereafter;
\citealt{Scargle1982ApJ...263..835S}).

In a first step, we searched for orbital variations.
Before our study, 8 of our targets were classified as confirmed or suspected
binaries (indicated with $^+$ in column (8) of Table\,\ref{targets}).
For \hd147787, \hd167858 and \hd209295 (indicated with $\ast$ in
Table\,\ref{elements}), an orbital solution was already available.
For all targets showing large Doppler shifts (typically $>$ 10~\kms) and/or
two components in the time series of CCFs, we tried to fit the large \vrad\
variations with an orbit by using an altered version of the {\sc fortran} code
{\sc vcurve} \citep{Bertiau1969}.

%%%%
\subsection{Solved binaries}
\label{solvedbin}
%%%%

%%%%
\subsubsection{Ellipsoidal variables}
%%%%

We took 2\phipp\ (with \phipp\ the main the photometric period as observed in
the \hipparcos\ variations; column (7) of Table\,\ref{targets}) as a first
trial for the orbital period.  
For the double-lined objects {\bf \hd34025} (Fig.\,\ref{hd34025fase}), 
{\bf \hd81421} (Fig.\,\ref{hd81421fase})\footnote{Figs.\,\ref{hd81421fase},
\ref{hd214291fase}, \ref{hd147787fase}, \ref{hd167858fase},
\ref{hd209295fase}, \ref{suspect}, \ref{hd41448fase}, \ref{hd112685fase},
\ref{hd187028fase}, \ref{hd125081fase}, \ref{hd135825fase},
\ref{hd218225fase}, \ref{hd149989fase}, \ref{hd216910fase},
\ref{CPVspreidingplot}, \ref{noCPVspreidingplot} are available only in
electronic form at the CDS via anonymous ftp to cdsarc.u-strasbg.fr
(130.79.128.5)},  
{\bf \hd214291} (Fig.\,\ref{hd214291fase})$^1$ and for the
single-lined object {\bf \hd85964} (Fig.\,\ref{hd85964fase}), 2\phipp\ fits
the observed \vrad\ variations.
We therefore classify them as ellipsoidal variables. 
According to the statistical test of \citet{Lucy1971AJ.....76..544L}, their
short-period orbits are all circular.  
For the determination of the final orbital elements (Table\,\ref{elements}),
the orbital period was fixed except for \hd214291 for which the total time
span of the \coralie\ data is almost 3 times larger than the total time span
of the \hipparcos\ data. 

\hd81421 was the only ellipsoidal variable in our sample which was already
classified as such \citep{Handler2002MNRAS.333..251H}.
\citet{Eyer2002oapb.conf..203E} classified both \hd81421 and \hd85964 as
eclipsing binaries because they observed a difference in depth of successive
minima in the \bvi\ light curves, while \citet{Duerbeck1997IBVS.4513....1D}
concluded from the period-colour-luminosity calibration of the \hipparcos\
data that \hd81421 is a contact binary.

\hd34025 and \hd214291 were previously classified as multi-periodic \gdor\
stars by \citet{Eyer2002oapb.conf..203E} and \citet{Aerts1998A&A...337..790A}
respectively. 
Since CPVs are observed for the primary component of \hd34025 (right panel of
Fig.\,\ref{hd34025fase}), we classify it as a bf \gdor\ star.
\hd81421 is the only ellipsoidal variable for which we have enough data to
search for additional intrinsic variations.
However, no CPVs are visible for either of its components, and there is no
evidence for intrinsic periods in the residual \vrad\ data. 
Hence, \hd81421 should be omitted from the list of cand \gdor\ stars.

%%%%
\subsubsection{Non-ellipsoidal variables}
%%%%

For 2 double-lined binaries, i.e. {\bf \hd10167} (Fig.\,\ref{hd10167fase}),
{\bf \hd147787} (Fig.\,\ref{hd147787fase})$^1$, and for 3 single-lined objects,
i.e. {\bf \hd126516} (Fig.\,\ref{hd126516fase}), {\bf \hd167858}
(Fig.\,\ref{hd167858fase})$^1$, {\bf \HD209295}
(Fig.\,\ref{hd209295fase})$^1$, \phipp\ is clearly not due to ellipsoidal
variations, but we were still able to determine the orbit (Table\,\ref{elements}).

\hd10167 was already reported as a double-lined object by
\citet{Grenier1999A&AS..135..503G} and by \citet{Eyer2000A&A...361..201E}. 
\citet{Handler2002MNRAS.333..251H} found in their photometric observations
only very little variations during individual nights, but night-to-night
variations of a few hundredths of a magnitude.
With our 15 \vrad\ values, we could determine the orbit for the first time
(Table\,\ref{elements}, column~2). 
There is no clear detection of CPVs in either of the components.
\hd147787 consists of two visual components ($V_A$\,= 5.29~mag; $V_B$\,=
9.12~mag) with a separation of 24.7~mas \citep{Landolt1969PASP...81..443L}. 
\hd147787 is also known as a single-lined spectroscopic binary with an
eccentric orbit of 37.4280~d \citep{Jones1928CA...10..68J}.
We here used the \vrad\ data of both components to determine orbits for a
grid of fixed eccentricity values of which we retained the one with the lowest
RMS value (Table\,\ref{elements}, column~7). 
Given the large errors on the results, this orbit should be fine-tuned with
additional spectroscopic observations.
For the determination of the orbits of \hd126515 and \hd167858, we included
the \vrad\ observations given by \citet{Fekel2003AJ....125.2196F}
and \citet{Fekel2003AJ....125.2156F} respectively. 
For \hd126516, it is the first time that an orbital solution is given
(Table\,\ref{elements}, column~6).
With our 7 \vrad\ observations of \hd167858, we can slightly improve the
orbit given by \citet{Fekel2003AJ....125.2156F} (cf. their Table\,2).   
For this object, several observations in speckle interferometry are available,
but no evidence of a companion was found so far
\citep{McAlister1987AJ.....93..183M,Mason1999AJ....117.1890M}. 
\hd209295 was first reported as a candidate spectroscopic binary by
\citet{Grenier1999A&AS..135..503G}.
The orbit given in column~9 of Table\,\ref{elements} is based on our 61
\vrad\ values. 
Our orbit is close to the orbit given by \citet{Handler2002MNRAS.333..262H}
(see their Table\,7), who already used a subset of our data amongst other
spectroscopic observations for this purpose. 

\hd147787, \hd167858 and \hd209295 are known as multi-periodic variables
(\citealt{Aerts1998A&A...337..790A}, \citealt{Handler2002MNRAS.333..251H},
\citealt{Fekel2003AJ....125.2156F}, \citealt{Handler1999MNRAS.309L..19H},
\citealt{Koen2001MNRAS.321...44K}, \citealt{Handler2002MNRAS.333..262H}, and/or \citealt{Eyer2002oapb.conf..203E}).  
For \hd167858, \citet{Mathias2004A&A...417..189M} already observed evident but
small line profile variations from their 11 \aurelie\ spectra, but a combined
fit of photometric and spectroscopic data was not possible. 
For \hd10167, no period is found in the \hipparcos\ photometry but it was
listed as a mono-periodic \gdor\ star by \citet{Eyer2002oapb.conf..203E}. 
\hd209295 is the only non-ellipsoidal binary for which we have enough data
to search for intrinsic variations in the residual \vrad\ data (see
Section\,\ref{intrinsic}).  
We classify the other ones as binaries with a (cand) \gdor\ component.
Because the further exploitation of their dynamical information with
e.g. interferometry can give additional and independent constraints on
physical properties of the components, these objects are the most interesting
ones in our sample from an asteroseismic point of view.

%%%%%%%%%%%
\subsection{Unsolved binaries}
\label{unsolvedbin}
%%%%%%%%%%%

\setcounter{figure}{10}

Finally, there is a group of 9 suspected binaries in our sample consisting
of systems for which the amount of data is insufficient to check if we are
dealing with an ellipsoidal variable or not and/or to determine an orbital
period. 
All their observed CCFs are shown in Fig.\,\ref{suspect}$^1$. 
At least 7 of these objects are double-lined.

\citet{Eyer2000A&A...361..201E} found no intrinsic photometric variations in
their \geneva\ data of {\bf \hd5590} and {\bf \hd8393}.
For {\bf \hd35416}, which is listed as ``microvariable'' in the {\sc
hipparcos} catalogue \citep{ESA1997}, they found a different result for each
period search method. However, their photometric observations of both \hd8393
and \hd35416 show a slight drift. 
These double-lined objects are clearly long period binaries: their \vrad\
values barely change within an observation run while clear Doppler shifts are
observed between observations of different observation runs.  
Hence, their orbital period must be much longer than 10~d.
Note that if we did not have the first observation of \hd35416
(Fig.\,\ref{suspect}, upper right panel, lowest CCF)$^1$, we would have
interpreted the CPVs as being due to pulsation instead of binarity.

{\bf \hd110606} is the only unsolved binary for which a composite
spectrum was observed before.
\citet{Nordstrom1997A&AS..126...21N} determined a mass ratio of 0.912(34).
It can not be an ellipsoidal variable since 2\phipp\ does not fit our \vrad\
data at all.
Our orbit determination failed, but the most promising candidate orbital
period is about 63~d. 

\citet{Paunzen1998A&AS..133....1P} concluded from the \hipparcos\ data that
{\bf \hd111709} is a new variable chemically peculiar star. 
The \bvi\ photometry of {\bf \hd27377} shows a long term behaviour that yields
a period of about 10~d \citep{Eyer2002oapb.conf..203E}. 
However, we suspect that it is a short period binary since we observed
$\Delta$ \vrad\,$\simeq$ 3~\kms\ in 45 minutes.
Our current spectroscopic data of these double-lined objects are insufficient
to confirm or rule out ellipsoidality.
Note that CPVs are present in the primary component of \hd111709
(Fig.\,\ref{suspect}, middle panel)$^1$, which confirms the \gdor\ or
spotted character of this component.  

In case of \hd27604, \hd26298 and \hd111829 (Fig.\,\ref{suspect}, bottom
panels)$^1$, it is not clear if we are dealing with binaries or not.
Indeed, \citet{Eyer2000A&A...361..201E} classified {\bf \hd27604} as a
constant star, but they observed a few data points with a higher than normal
magnitude, which might indicate the presence of eclipses. 
It is a close visual binary with a separation of 0.800~arcsec \citep{ESA1997},
which might explain why we observe a double-lined object but without clear
Doppler shifts. 
{\bf \hd26298} is listed as a constant star in the {\sc hipparcos} catalogue
\citep{ESA1997}.
\citet{Lu1987AJ.....94.1318L} found no evidence for duplicity with speckle
interferometry. 
The \geneva\ observations of \citet{Eyer2000A&A...361..201E} yield periods
close to a sampling period while the period in colour seems to be half the
period in $V$, which points towards ellipsoidal variability.
We find a peak-to-peak value of 9.5~\kms\ in \vrad.
The CCFs do not show clear asymmetries, but rather global Doppler shifts, which
favours an interpretation as a suspected binary. 
For {\bf \hd111829}, \citet{Mantegazza1991IBVS.3612....1M} observed not
strictly periodic light variations with a $V$ amplitude of about 0.04~mag on a
time-scale of 1.8 or 4~d.
%\citet{Handler1995IBVS.4216....1H} notes that this suspected \gdor\ variable
%is hotter and more evolved than confirmed \gdor\ variables.
Our 2 observations show $\Delta$ \vrad\,$\simeq$ 7~\kms\ in 1 hour.
It is unclear if the obvious line profiles observed in the CCFs are superposed
on line profiles of a very rapidly rotating component or not
(Fig.\,\ref{suspect}, lower right panel)$^1$.  

\smallskip
Two of these unsolved binaries were previously classified as multi-periodic
variables, i.e. \hd27377 by \citet{Handler1999MNRAS.309L..19H}, and \hd110606
by \citet{Eyer2002oapb.conf..203E}. 
Since we cannot derive their orbits, we are unable to search for intrinsic
periods. 
Moreover, apart from the primary of \hd111709, none of the components of the
unsolved binaries shows clear CPVs.

\section{Intrinsic variations}
\label{intrinsic}

Before our study, 6 of our targets were classified as bf \gdor\ stars
(indicated with $\degree$ in column (10) of Table\,\ref{targets}). 
In the second step, we searched for intrinsic periods in the
(residual) \vrad\ data of all the objects with clear CPVs as far as the number
of data points allowed it. 
In most cases, the resulting periodograms are very noisy. 
If our data turned out to be insufficient for an independent period search, we
made phase diagrams with already known photometric periods to check the
possible consistency with the observed spectroscopic variations.  

%----------
\subsection{Clear correlation profile variations}
\label{CPV}
%----------

\begin{figure*}
\begin{center}
\resizebox{6.70cm}{!}{\rotatebox{270}{\includegraphics{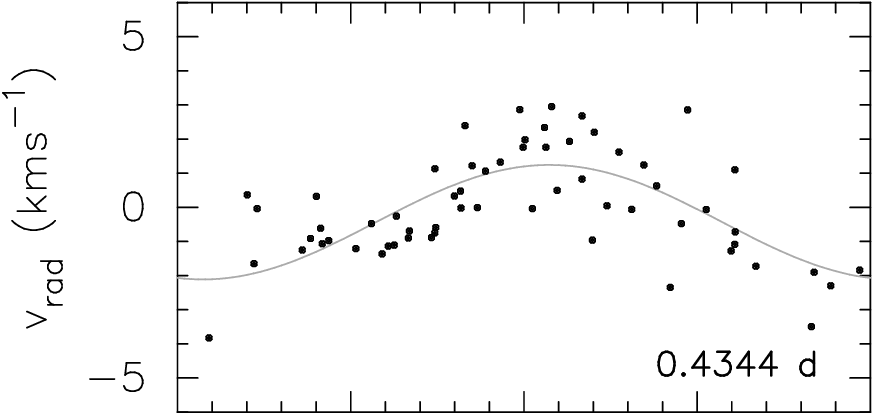}}}
\resizebox{6.03cm}{!}{\rotatebox{270}{\includegraphics{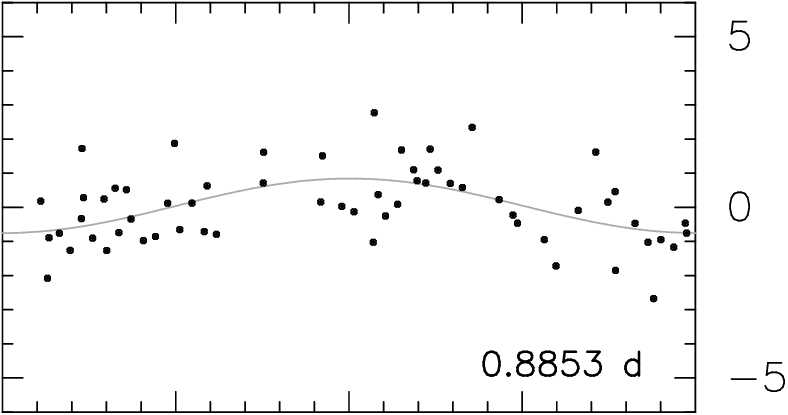}}}\\
\resizebox{6.70cm}{!}{\rotatebox{270}{\includegraphics{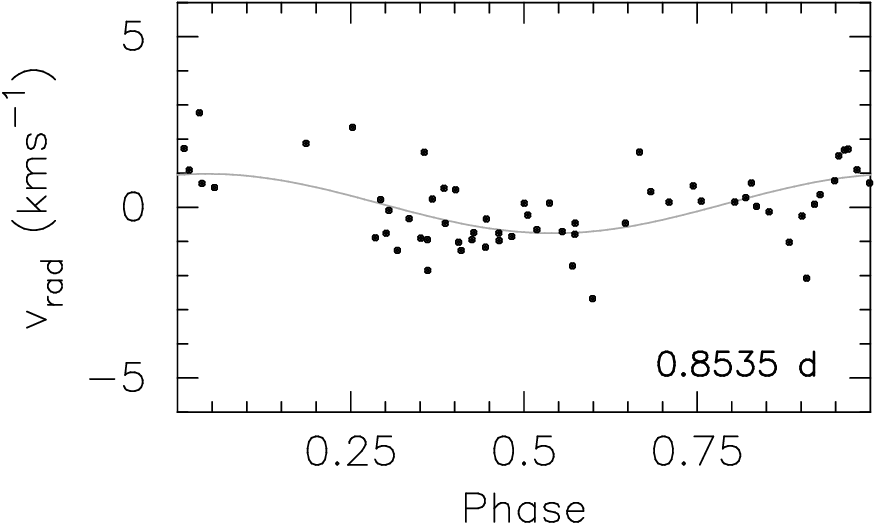}}}
\resizebox{6.03cm}{!}{\rotatebox{270}{\includegraphics{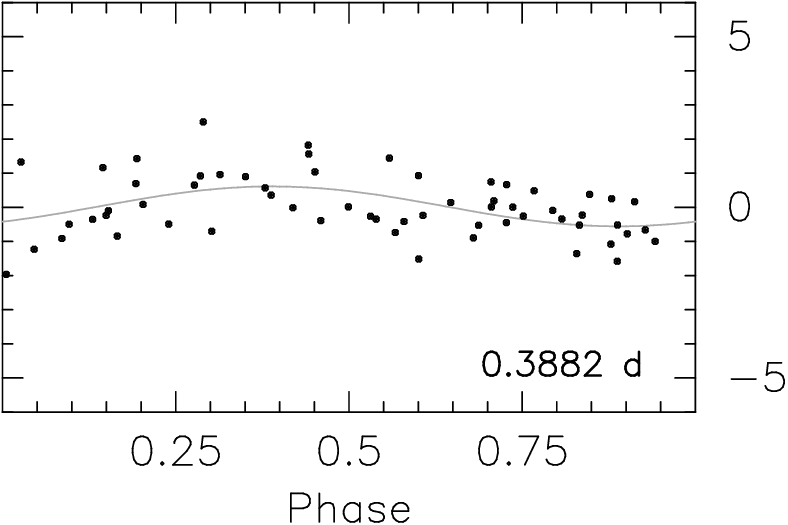}}}
\caption{\label{hd209295faseres} 
Phase diagrams for the intrinsic periods of \hd209295 found in the residual
\vrad\ data after prewhitening with the orbit given in Table\,\ref{elements}.
The reference epoch is HJD 2450000.
Top left: \pa\,= 0.4343(2)~d. Top right: \pb\,= 0.8853(9)~d\
(prewhitened with \pa). Bottom left: \pc\,= 0.8535(9)~d\ (prewhitened with
\pa). Bottom right: \pd\,= 0.3882(2)~d\ (prewhitened with \pa, \pb\ and
\pc). 
}
\end{center}
\end{figure*}

\begin{figure*}
\begin{center}
\resizebox{7.42cm}{!}{\rotatebox{270}{\includegraphics{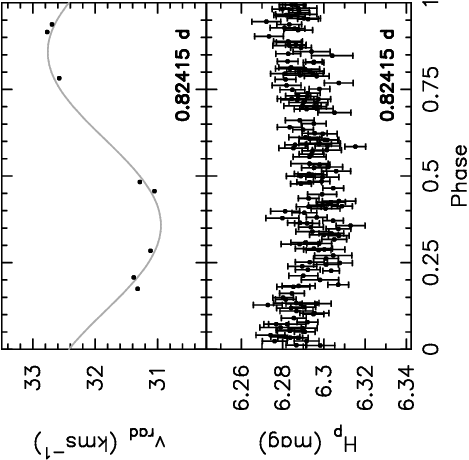}}}
\resizebox{5.20cm}{!}{\rotatebox{270}{\includegraphics{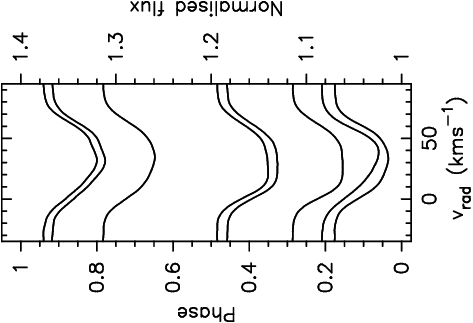}}}
\end{center}
\caption{\label{hd40745fase} 
Phase diagram of the radial velocity \vrad\ (top left) and the Hipparcos \hp\
measurements (bottom left) of \hd40745 with the period as given in the bottom
right corner. 
The reference epoch is HJD 2450000.
In the right panel, a selection of observed cross-correlation profiles are
shown as a function of pulsation phase.
}
\end{figure*}

\setcounter{figure}{13}

\setcounter{figure}{14}

\setcounter{figure}{15}

\setcounter{figure}{16}

\setcounter{figure}{17}

\setcounter{figure}{18}

\begin{figure*}
\begin{center}
\resizebox{7.42cm}{!}{\rotatebox{270}{\includegraphics{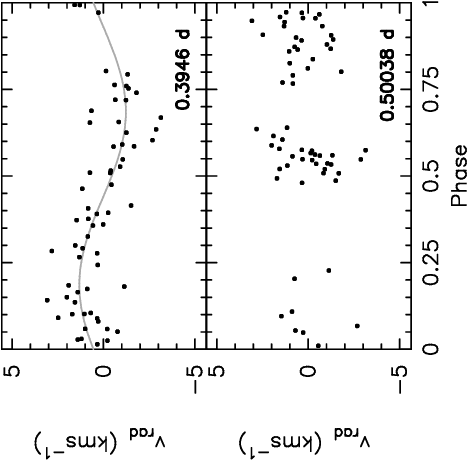}}}
\resizebox{5.20cm}{!}{\rotatebox{270}{\includegraphics{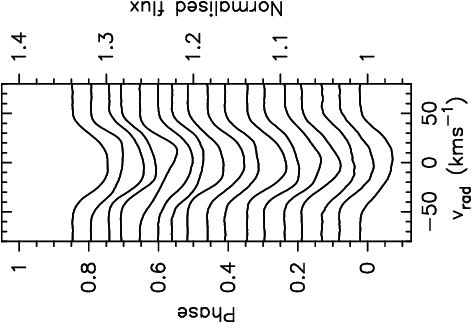}}}
\end{center}
\caption{\label{hd14940fase} 
Phase diagram of the radial velocity \vrad\ with our best fitting
period (top left) and the best fitting photometric period
(bottom left) of \hd14940. 
The period values are given in the bottom right corner.
The reference epoch is HJD 2450000.
In the right panel, a selection of observed cross-correlation profiles are
shown as a function of our best fitting period. 
}
\end{figure*}

\begin{figure*}
\begin{center}
\resizebox{7.42cm}{!}{\rotatebox{270}{\includegraphics{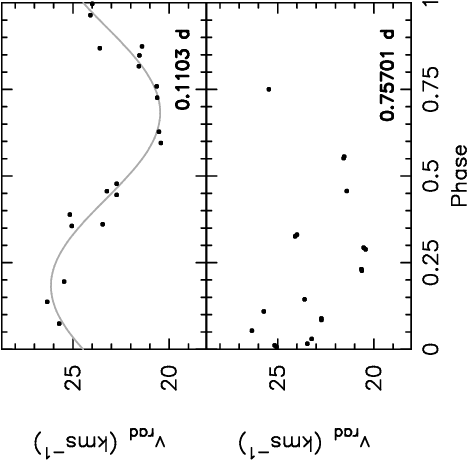}}}
\resizebox{5.20cm}{!}{\rotatebox{270}{\includegraphics{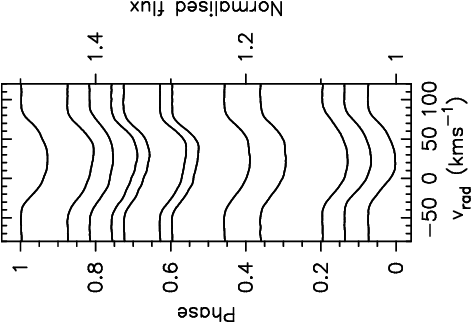}}}
\end{center}
\caption{\label{hd27290fase} 
Same as Fig.\,\ref{hd14940fase}, but for \hd27290.
}
\end{figure*}

%%%%
\subsubsection{Spectroscopic variations compatible with photometric period}
%%%%

\hd34025, \hd111709 and \hd209295 are the only binaries in our sample for
which clear CPVs are seen in one of the components (right panels of
Figs.\,\ref{hd34025fase} and \ref{hd209295fase}).
Only for {\bf \hd209295}, the number of datapoints and the quality of the
derived orbit allow us to search for additional intrinsic periods.
\citet{Handler2002MNRAS.333..262H} organised a multi-site campaign for this
object and detected multi-periodic \gdor\ and \dsct\ variability.
A total of 10 independent periods were found: 7 in photometry
(\pap\,= 0.88529(2)~d, \pbp\,= 0.434363(6)~d, \pcp\,= 0.388230(8)~d,
\pdp\,= 0.8519~d, \pep\,= 0.5659~d, \pfp\,= 0.4430~d, \pgp\,= 0.9246~d) and 3
additional ones in spectroscopy (\phs\,= 1.03523(12)~d, \pis\,= 0.62113(4)~d,
\pjs\,= 0.345075(13)~d).
\pcp, \pfp, \phs, \pis\ and \pjs\ are suspected to be tidally excited
since they are exact subharmonics of the orbital period.
We have evidence that at least 4 of the known periods are also present in the
residual \vrad\ data after prewhitening the orbit given
in Table\,\ref{elements}: \pa\,= 0.4343(2)~d\,$\simeq$ \pbp, \pb\,= 
0.8853(9)~d\,$\simeq$ \pap, \pc\,= 0.8535(9)~d\,$\simeq$ \pdp, and \pd\,=
0.3882(2)~d\,$\simeq$ \pcp\ (Fig.\,\ref{hd209295faseres}).
During the period search, 0.8650(9)~d originally seemed to be the best
candidate for the second period. 
However, since (1/0.8650)\,$\simeq$ (1/\pap\,+1/\pdp)/2, and since prewhitening
with 0.8650~\cd\ or with \pb\ and \pc\ reduces the variance with the same
amount, we favoured the latter two.
We refer to \citet{Handler2002MNRAS.333..262H} for a more extensive study of
this star.

For the objects {\bf \hd40745} (Fig.\,\ref{hd40745fase}), {\bf \hd41448}
(Fig.\,\ref{hd41448fase})$^1$, {\bf \hd112685} (Fig.\,\ref{hd112685fase})$^1$ and 
{\bf \hd187028} (Fig.\,\ref{hd187028fase})$^1$, a period close to \phipp\ (or one
of its aliases) is found as one of the best fitting periods from an
independent search in our \vrad\ data.
We therefore have evidence that the main photometric and spectroscopic periods
coincide for these objects. 
These periods are not connected to binarity, and, since the \vrad\ variations
are sinusoidal, %and have a very low velocity amplitude ($<$ 5~\kms), 
they are probably not due to spots.
Moreover, \hd40745, \hd41448 and \hd187028 show multi-periodic variations in
their \hipparcos\ observations (e.g. \citealt{Handler1999MNRAS.309L..19H})
while \hd112685 was found to be multi-periodic in \bvi\ observations
\citep{Eyer2002oapb.conf..203E}.
We can therefore classify them as bf \gdor\ stars.
Note that \citet{Mathias2004A&A...417..189M} failed to observe line profile
variations in their \aurelie\ data of \hd40745 and \hd41448.
In case of \hd40745, we can not confirm any of the alternative \hipparcos\
periods given by \citet{Aerts1998A&A...337..790A}. 

The period \phipp\,= 0.15398~d observed in the \hipparcos\ observations of {\bf
\hd125081} \citep{ESA1997} is typical for \dsct\ stars.
\citet{Paunzen1998A&AS..133....1P} classified this object as a new variable
chemically peculiar star (F3\,SrCrEu). 
Our observations confirm the \dsct\ character of \hd125081, since \phipp\ is
clearly the main period in spectroscopy (Fig.\,\ref{hd125081fase})$^1$

For {\bf \hd135825} and {\bf \hd218225}, none of the {\it highest} peaks in
the periodograms correspond to a period close to \phipp\ or one of its
aliases. 
The best fitting periods, 0.63(3)~d and 2.0210(4)~d respectively, reduce the
variance with more than 80\%. 
However, for both objects, {\it low} peaks do occur near \phipp\ in the
periodograms, and these periods already reduce the variance with $\sim$ 60\%.
Hence, given the small number of datapoints, we can not rule out that \phipp\
coincides with the main spectroscopic period.
In the left panels of Figs.\,\ref{hd135825fase}$^1$ and
\ref{hd218225fase}$^1$, phase diagrams with \phipp\ are given.
Since both \hd135825 and \hd218225 are multi-periodic in photometry
\citep{Eyer2002oapb.conf..203E,Aerts1998A&A...337..790A}, we now
classify them as bf \gdor\ stars.

%%%%
\subsubsection{Spectroscopic variations not compatible with photometric period}
%%%%

For {\bf \hd14940}, two periods were detected in the \hipparcos\ data: 0.5004~d
and 0.9800~d \citep{Aerts1998A&A...337..790A}.
However, they do not fit our \vrad\ data at all.
These periods reduce the variance with less than 10\% while the best fitting period,
0.3946~d, induces a variance reduction of more than 50\%
(Fig.\,\ref{hd14940fase}).  
After prewhitening, there is still power near 0.5842~d, 0.9647~d and 0.4234~d.
We are clearly dealing with a multi-periodic \gdor\ star, but the aliasing is too
strong to continue our period search. 

For \gdor\ (= {\bf \hd27290}), three periods are commonly present in
photometry and spectroscopy: 0.75701~d, 0.73339~d and 0.67797~d
\citep{Balona1994MNRAS.267..103B, Balona1994MNRAS.270..905B,
Balona1996MNRAS.281.1315B}.  
However, we find a \dsct\ like period instead: 0.1103~d reduces the variance
with 90\% while the variance reduction for the known periods are all below
45\% (Fig.\,\ref{hd27290fase}).
We do stress that the periodograms are very noisy, and we clearly need more data
to confirm or rule out this result. 

\setcounter{figure}{21}

\setcounter{figure}{22}

\citet{Aerts1998A&A...337..790A} found two periods in the \hipparcos\ data of 
both {\bf \hd149989} and {\bf \hd216910}.
In case of \hd216910, none of these periods could be confirmed with the
\bvi\ observations by \citet{Eyer2002oapb.conf..203E}. 
Our current amount of \coralie\ data is insufficient for an independent period
search.  
However, phase plots with the \hipparcos\ periods enable us to rule out their
presence in our spectroscopic observations
(Fig.\,\ref{hd149989fase} and Fig.\,\ref{hd216910fase})$^1$. 
Due to the broad CCPs, the CPVs of \hd149989 are not as clear as for the
objects discussed in Section\,\ref{CPV} so far.

%%%%
\subsubsection{Unsolved intrinsic variations}
%%%%

For {\bf \hd65526}, the \hipparcos\ team found \phipp\,= 1.28798~days as the main
period \citep{ESA1997}.
A total of four other periods were derived with the \hipparcos\ data by
\citet{Handler1999MNRAS.309L..19H} and by \citet{Martin2003A&A...401.1077M},
who confirmed one of them in their Str\"omgren data.
\citet{Mathias2004A&A...417..189M} did not detect asymmetries in their 1
\aurelie\ spectrum, while the CPVs are obvious our 2 CCFs
(Fig.\,\ref{CPVspreidingplot})$^1$.  

\setcounter{figure}{23}

%----------
\subsection{No clear correlation profile variations}
\label{noCPV}
%----------

\setcounter{figure}{24}

\begin{table}
\caption{\label{sigma} 
The peak-to-peak value ($\Delta v_{\rm rad}$) and the standard deviation
($\sigma _{\rm rad}$) of the \vrad\ data is given for the objects for which no
clear correlation profile variations have been detected in our current set of
\coralie\ data.   
}
\begin{center}
\begin{tabular}{lclllll} \hline 
& \\ [-8pt]
HD          & HIP     & $\Delta v_{\rm rad}$ & $\sigma _{\rm rad}$ \\ [2pt] \hline \\ [-8pt]
 22001      &  16245  &  0.2 &  0.05 &     & const \\
 33262      &  23693  &  0.2 &  0.06 &     & const \\
  7455      &   5745  &  0.3 &  0.07 &     & const \\ [2pt]
112934$\ast$&  63491  &  1.2 &  0.5  &     & CPVs?  \\
110379      &         &  2.5 &  0.8  &     & CPVs? \\  [2pt] \hline
\end{tabular}
\end{center}
\end{table}

There are five (apparently) single objects for which no clear CPVs are detected
in our \coralie\ data (Fig.\,\ref{noCPVspreidingplot})$^1$.
{\bf \hd7455} and {\bf \hd22001} were found to be constant in \geneva\
photometry \citep{Eyer2000A&A...361..201E}.
For {\bf \hd33262}, a period of 3.4961~d was found in the \hipparcos\ data
\citep{Koen2002MNRAS.331...45K} while only some indications of a period of 
about 100 days was found in \geneva\ data \citep{Eyer2000A&A...361..201E}.
For these three objects, we find peak-to-peak values $\Delta v_{\rm rad}$
below 1~\kms, and corresponding standard deviations $\sigma _{\rm rad}$ well
below 0.1~\kms\ (Table\,\ref{sigma}). 
Hence classify them as spectroscopically constant. 

In the \hipparcos\ data of {\bf \hd112934}, \citet{Handler1999MNRAS.309L..19H}
found evidence for a period of $\sim$ 0.8~days.
\citet{Eyer2002oapb.conf..203E} confirmed its classification as a mono-periodic
\gdor\ star on the basis of new \bvi\ observations. 
For {\bf \hd110379}, no \hipparcos\ observations are available, but
\citet{Krisciunas1995IBVS.4195....1K} observed a period of 0.228~days in
Str\"omgren photometry. 
With our \coralie\ data alone, it is not clear if we are dealing with
pulsating stars or not. 
No clear CPVs are observed (Fig.\,\ref{noCPVspreidingplot})$^1$, but their
$\sigma _{\rm rad}$ values are at least 10 times larger than the one of \hd22001.

\section{Projected rotational velocity}
\label{methods}

\begin{figure*}
\begin{center}
\resizebox{12.50cm}{!}{\rotatebox{270}{\includegraphics{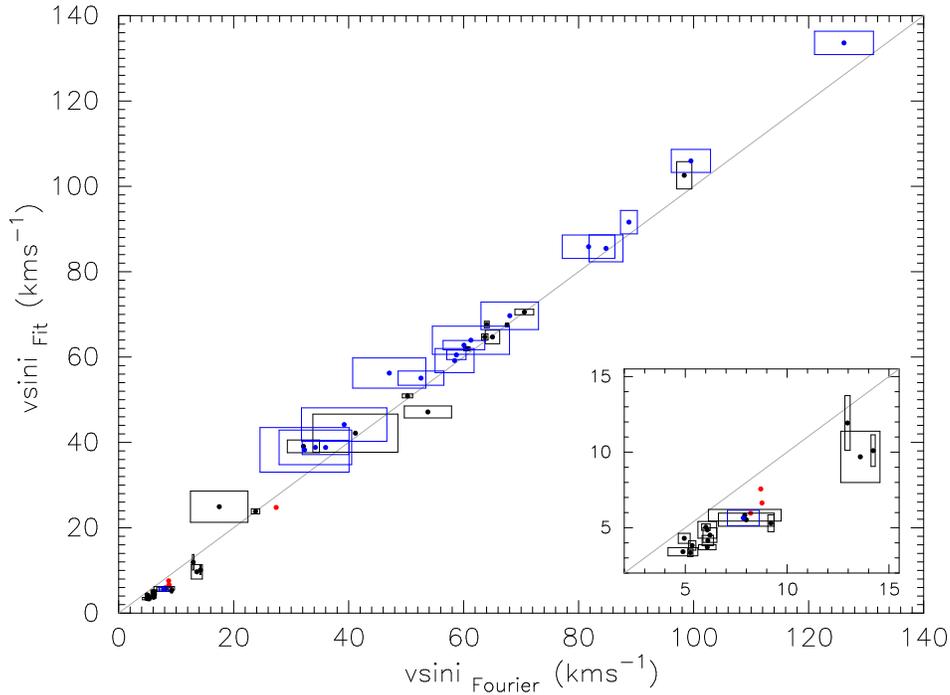}}}
\end{center}
\caption{\label{vsini} 
Comparison of the \vsini\ values (full circles with error boxes) determined
with the Fourier method (\vsinifou) to those determined by least-squares
fitting with synthetic profiles (\vsinifit). 
%The bisector is indicated with a grey line.
The components known to be pulsating are given in blue and those for which the
\vsini\ value is based on 1 cross-correlation profile only are given in red.
Colour representation only in the online verions of the paper.
}
\end{figure*}

The CCFs were used to determine the projected rotational velocity (\vsini). 
For the double-lined objects, only the CCFs in which the profiles of both
components are well separated are used (if available).
The \vsini\ values were determined with the Fourier method
(\vsinifou; \citealt{Gray1992}) and by least-squares fitting with rotationally
broadened synthetic profiles with a Gaussian intrinsic width but without
pulsational broadening (\vsinifit).
For each object, a value of 0.555 was taken for the limb-darkening
coefficient.

In Fig.\,\ref{vsini}, the resulting \vsini\ values and their corresponding
standard deviations as found with both methods are compared for each
component. 
%The \vsini\ values range from 3 to 135 \kms. 
For \vsini\,$>$ 15~\kms, the results of both methods are generally fully
compatible, while for \vsini\,$<$ 15~\kms\ (see enlargement in Fig.\,\ref{vsini}), we systematically find
\vsinifou\,$>$ \vsinifit. 
For slow rotators, there is confusion in the zeros of the
Fourier transform invoked by broadening mechanisms other than rotation.
Moreover, the standard deviations on the \vsini\ values found with the Fourier
method are in general larger than those found with least-squares fitting. 
We therefore favour the results of the least-squares fitting and list the
\vsinifit\ values for the primary and secondary components in columns (11) and
(12) of Table\,\ref{targets} respectively. 
Although the cross-correlation profiles of pulsating components show
asymmetries, we do not detect significant differences in the relative errors
on \vsini\ of the pulsating and non-pulsating components in our sample,
given in blue and black respectively in Fig.\,\ref{vsini}\footnote{The colours
are given in the electronic version only}.

\begin{figure}
\begin{center}
\resizebox{7.0cm}{!}{\rotatebox{270}{\includegraphics{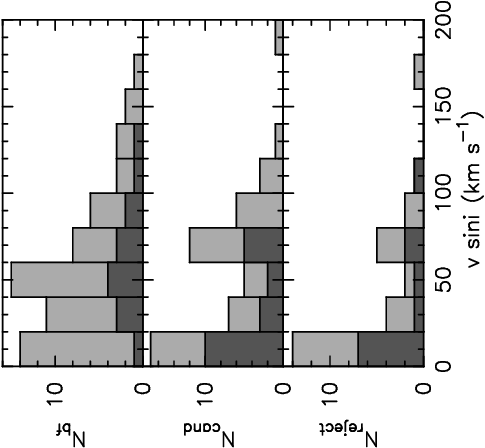}}}
\end{center}
\caption{\label{convsini} 
Histograms showing the number of bf \gdor\ stars (N$_{\rm bf}$;
top), cand \gdor\ stars (N$_{\rm cand}$; middle) and rejected \gdor\
stars which were formerly under consideration (N$_{\rm reject}$; bottom) as a
function of \vsini.
The \vsini\ values from this work are given in dark grey.
The other values were taken from \citet{Handler2002,Henry2002PASP..114..988H,
Henry2003AJ....126.3058H,Fekel2003AJ....125.2196F,
Mathias2004A&A...417..189M,Henry2005AJ....129.2026H,Henry2005AJ....129.2815H}. 
}
\end{figure}

In Fig.\,\ref{convsini}, we show histograms of the number of bf \gdor\
stars (N$_{\rm bf}$), cand \gdor\ stars (N$_{\rm cand}$) and rejected
\gdor\ stars (i.e. objects which were formerly under consideration; N$_{\rm
  reject}$), as a function of \vsini. 
As input for the different groups, we used a compilation of our results and
those listed by \citet{Handler2002}, \citet{Henry2002PASP..114..988H},
\citet{Henry2003AJ....126.3058H}, \citet{Fekel2003AJ....125.2196F},
\citet{Mathias2004A&A...417..189M}, \citet{Henry2005AJ....129.2026H}, and
\citet{Henry2005AJ....129.2815H}. 
One third of the shown \vsini\ values used were determined in this work (dark grey). 
In case of a double-lined object for which (only) one of the components is
classified as a bf \gdor\ star, the companion was considered as a
rejected \gdor\ star.
If a double-lined object is classified as a cand/rejected \gdor\ star,
both components were considered as such.  
The histogram of the bf \gdor\ stars shows an excess of objects within
\vsini\,$\in$ $[20,60]$~\kms\ compared to the cand and rejected \gdor\ stars. 
These rotation rates are indeed ideal to detect profile variations: very
slow rotators require a very high spectral resolution for a detailed view of
the line profile while rapid rotators suffer from line blending. 
There is no evidence for pulsation damping due to rapid rotation
because the relative number of stars with a high \vsini\ value is comparable
in the 3 categories.

\section{Conclusions \& Future prospects}
\label{discussion}

\begin{table*}
\caption{\label{summary} 
An overview of the orbital classification, i.e. apparently single star
(``single''), suspected binary (``suspect''), single-lined spectroscopic
binary (``SB1'') and double-lined spectroscopic binary (``SB2''), and
variability classification of the objects discussed in this paper.
The stars in {\it italics} will be subject of dedicated future studies.
}
\begin{center}
\begin{tabular}{|l|ll|ll|ll|} \hline 
        &    &                                        &    &                                            &    &                                   \\[-8pt]
        &\multicolumn{2}{|l|}{bf \gdor\ star}  &\multicolumn{2}{|l|}{cand \gdor\ star}      &\multicolumn{2}{|l|}{rejected \gdor\ star}\\[2pt] \hline 
        &    &                                        &    &                                            &    &                                   \\[-8pt]
single  & 13 &{      \hd12901, \hd14940, \hd27290,   }&  2 &{      \hd110379, \hd112934                }&  4 &{      \hd7455, }\\
        &    &{      \hd40745, \hd41448, \hd48501,   }&    &{                                          }&    &{      \hd22001,}\\
        &    &{      \hd65526, \hd112685, \hd135825, }&    &{                                          }&    &{      \hd33262,  }\\
        &    &{      \hd149989, \hd187025, \hd216910,}&    &{                                          }&    &{      \hd125081$^1$          }\\
        &    &{      \hd218225                       }&    &{                                          }&    &{               }\\[2pt] \hline 
        &    &                                        &    &                                            &    &                 \\[-8pt]
suspect &  0 &{                                      }&  2 &{\it   \hd111829, \hd26298                 }&  1 &{      \hd27604 }\\[2pt] \hline
        &    &                                        &    &                                            &    &                 \\[-8pt]
SB1     &  2 &{\it   \hd167858$^2$, \hd209295        }&  1 &{\it   \hd126516                           }&  1 &{      \hd85964 }\\[2pt] \hline 
        &    &                                        &    &                                            &    &                 \\[-8pt]
SB2     &  1 &{\it   \hd34025                        }&  7 &{\it   \hd10167, \hd27377$^3$, \hd35416    }&  3 &{      \hd5590, }\\ 
        &    &{                                      }&    &{\it   \hd110606, \hd111709$^{3,4}$, \hd147787,}&&{      \hd8393, }\\
        &    &{                                      }&    &{\it   \hd214291                           }&    &{      \hd81421 }\\[2pt] \hline 
\end{tabular}\\
$^1$ bf \dsct\ star; 
$^2$ shows no cross-correlation profile variations but was classified as bf \gdor\ star before;\\ 
$^3$ ellipsoidal variability instead of pulsation can not be ruled out;\\
$^4$ shows cross-correlation profile variations but was classified as chemically peculiar star before
\end{center}
\end{table*}

We contributed to the observational effort to confirm cand \gdor\ stars as
real members of the group of \gdor\ stars by analysing the time-series of
the \coralie\ spectra of 37 {\it southern} (cand) \gdor\ stars.
To allow a better detection of secondary components and/or profile variations,
the original spectra were cross-correlated with the standard template spectrum of
an F0-type star.
An overview of our final results is given in Tables\,\ref{targets} and
\ref{elements}, and of our binarity and variability classification in
Table\,\ref{summary}.  

At least 15 of our 37 targets turn out to be spectroscopic binaries, including
7 new ones.
Our data allowed the determination of 9 orbits, of which 6
are new (Table\,\ref{elements}): \hd34025, \hd81421, \hd214291 and \hd85964
are ellipsoidal variables, and \hd10167, \hd126516, \hd167858, \hd147787 and
\hd209295 are binaries with a (cand) \gdor\ component. 
For \hd34025, \hd81421 and \hd85964, the phases at which \vrad\,= $v_{\gamma}$
do not correspond to the phases of minimal light seen in \hipparcos\ photometry.
For the remaining 6 binaries, we estimate from our data an orbital period
\porb\,$>>$ 10~days for \hd5590, \hd8393, \hd35416 and \hd110606, and \porb\
of a few days for \hd27377 and \hd111709. 
For the latter two objects, we can not exclude ellipsoidal variations.
We additionally classify \hd27604, \hd26298 and \hd111829 as candidate
binaries.
Especially in case of \hd26298, it is not clear if the observed variations in
the CCFs are caused by binarity or not.
We find a binarity rate of about 50\% for our sample, which is similar
to the one found by \citet{Mathias2004A&A...417..189M}. 
If we restrict our sample to the objects which we classified as bf or
cand \gdor\ stars (see column (10) in Table\,\ref{targets}), the binary
rate lowers to $\sim$40\%. 
At least 12 of our objects, i.e. $\sim$~1/3 of our targets, show
composite spectra. 

For 17 objects ($\sim$45\%), including 3 binaries, clear
cross-correlation profile variations are observed in one of the components.
We classify them as bf \gdor\ stars, except for \hd125081 (which is a
bf \dsct\ star) and \hd111709 (for which there is possible confusion with
chemical peculiarity).
This confirms the \gdor\ character of 10 objects (\hd14940, \hd34025, \hd40745,
\hd41448, \hd112685, \hd135825, \hd149989, \hd187028, \hd216910 and \hd218225)
and the \dsct\ character of 1 object (\hd125081).
For 8 objects, we have evidence that the intrinsic spectroscopic variations
are compatible with (one of) the known photometric period(s).
For 4 others, we observe no such compatibility.
This latter group includes \gdor\ itself (\hd27290), for which our data points
towards a \dsct\ period instead. 
Due to a lack of evidence for intrinsic spectroscopic and photometric
variations, we classify \hd5590, \hd7455, \hd8393, \hd22001, \hd27604,
\hd33262, \hd81421 and \hd85964 as constant stars.
The variability classification of our other objects remains unchanged.

The cross-correlation profiles were used to
determine accurate values for the projected rotational velocity \vsini\ with 2
independent methods.
The resulting values range from 3 to 135~\kms.
The \vsini\ values gathered for all the bf, cand and rejected
\gdor\ stars result into decreasing histograms towards high \vsini\ values,
except for the bf \gdor\ stars which show more stars in the range of
$[20,60]$~\kms. 

%{\bf synchronisation of double-lined objects}
% synchronisatie komt voor circularisatie
% in staat van minimale energie zijn zowel rotatie-orbitale periodes als de
% rotatie-orbitale inclinaties hetzelfde
% circularisatie werkt efficient bij een convectieve enveloppe, dus is zeer
% efficient in de pre-main sequence fase, maar door de contractie van de
% proto-sterren worden de sterren kleiner waardoor ze sneller gaan draaien, en
% daardoor kan het zijn dat de synchronisatie opgeheven wordt terwijl de baan
% al wel gecirculariseerd was.
Tidal evolution of binary systems leads to circularisation of the orbit,
spin-alignment of the rotational and orbital inclination, and synchronisation
of the rotation of the components with the orbital motion.  
For 5 double-lined binaries in our survey, i.e. \hd8393, \hd27377, \hd10167,
\hd34025 and \hd214291, the \vsini\ values of both components are (close to)
equal, which reflects their spin-alignment and synchronisation. 
Moreover, the orbits of \hd10167, \hd34025 and \hd214291 are circularised
(Table\,\ref{elements}). 
For the other double-lined {\it binaries} in our sample, i.e. \hd5590,
\hd34025, \hd81421, \hd110606, \hd111709 and \hd147787, the evolution towards
synchronisation and/or spin-alignment is still ongoing.
However, the orbits of \hd34025 and \hd81421 are already circularised
(Table\,\ref{elements}).  

The binarity and variability classifications of each of our target stars were
updated by the use of 2 up to 63 \coralie\ spectra observed from a single site
(Table\,\ref{summary}). 
The low number of observations and the strong aliasing made an individual
orbital or intrinsic period search difficult or impossible in several cases.
Therefore, a lot of our results are based on a comparison with photometric
observations of the satellite mission \hipparcos. 
Clearly, additional (and ideally {\it multi-site}) spectroscopic data are
needed to determine or improve orbits, and to study the intrinsic variations
present in spectroscopic observations in more detail.
Because the exploitation of dynamical information can give additional
and independent constraints on physical properties of the components,
we will give priority to binaries with a bf or cand \gdor\ star in
future investigations.
These objects are given in {\it italics} in Table\,\ref{summary}.

\acknowledgements{This research was made possible thanks to the financial
support from the Fund for Scientific Research - Flanders (FWO), under project
G.0178.02 and from the Research Council of the University of Leuven under grant
GOA/2003/04.
The authors performed their work within the Belgian Asteroseismology Group (BAG, http://www.asteroseismology.be/).  
This research has made use of the SIMBAD astronomical database operated at the
CDS in Strasbourg, France.
We are grateful for the valuable suggestions and remarks from our referee,
Dr. P. Mathias, which have improved this manuscript.}

\newpage

{\bf \huge ONLINE MATERIAL}

\setcounter{figure}{1}

\begin{figure*}[h]
\begin{center}
\resizebox{7.42cm}{!}{\rotatebox{270}{\includegraphics{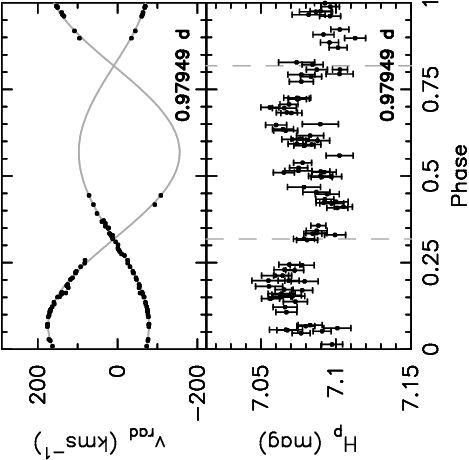}}}
\resizebox{5.20cm}{!}{\rotatebox{270}{\includegraphics{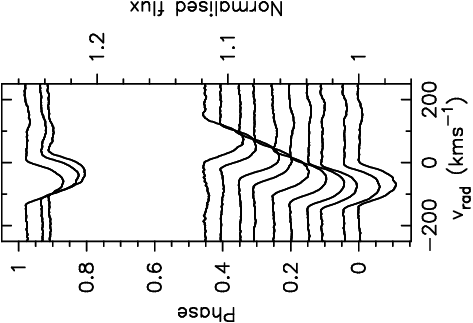}}}
\end{center}
\caption{\label{hd81421fase} 
Same as Fig.\,\ref{hd34025fase}, but for \hd81421.
}
\end{figure*}

\begin{figure*}[h]
\begin{center}
\resizebox{7.42cm}{!}{\rotatebox{270}{\includegraphics{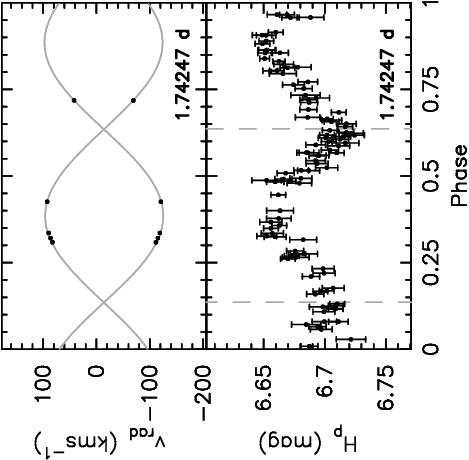}}}
\resizebox{5.20cm}{!}{\rotatebox{270}{\includegraphics{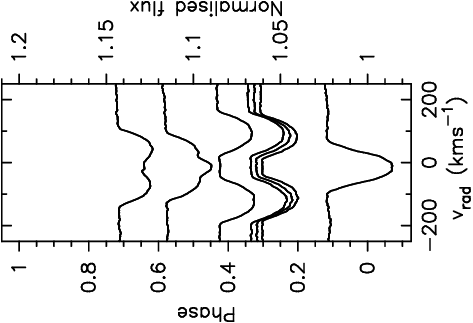}}}
\end{center}
\caption{\label{hd214291fase} 
Same as Fig.\,\ref{hd34025fase}, but for \hd214291.
}
\end{figure*}

\setcounter{figure}{5}

\begin{figure*}[h]
\begin{center}
\resizebox{7.42cm}{!}{\rotatebox{270}{\includegraphics{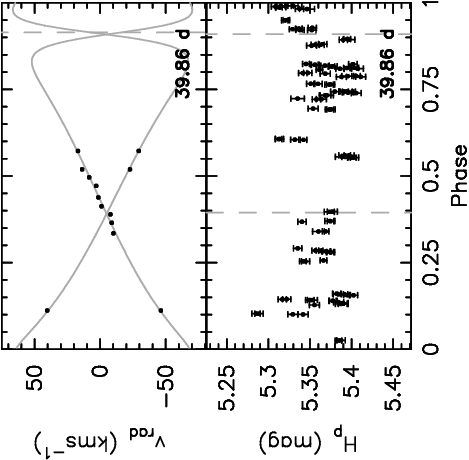}}}
\resizebox{5.20cm}{!}{\rotatebox{270}{\includegraphics{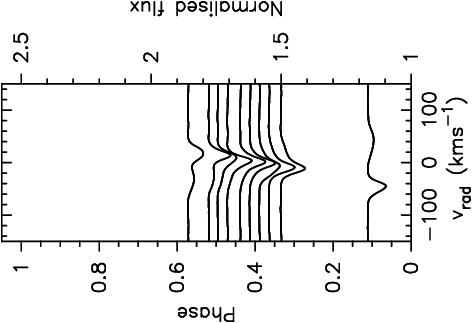}}}
\end{center}
\caption{\label{hd147787fase} 
Same as Fig.\,\ref{hd34025fase}, but for \hd147787.
The dotted line in the top left panel indicates the time of periastron passage.
}
\end{figure*}

\setcounter{figure}{7}

\begin{figure*}[h]
\begin{center}
\resizebox{7.42cm}{!}{\rotatebox{270}{\includegraphics{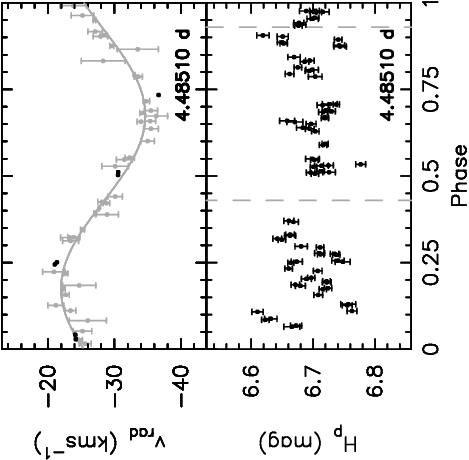}}}
\resizebox{5.20cm}{!}{\rotatebox{270}{\includegraphics{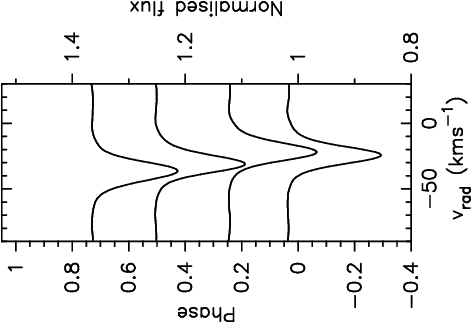}}}
\end{center}
\caption{\label{hd167858fase} 
Same as Fig.\,\ref{hd34025fase}, but for \hd167858.
In the top panel, the radial velocities, as determined by
\citet{Fekel2003AJ....125.2156F} are included and given in grey.
}
\end{figure*}

\begin{figure*}[h]
\begin{center}
\resizebox{7.42cm}{!}{\rotatebox{270}{\includegraphics{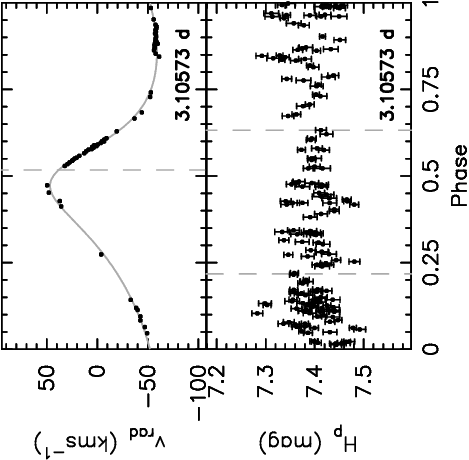}}}
\resizebox{5.20cm}{!}{\rotatebox{270}{\includegraphics{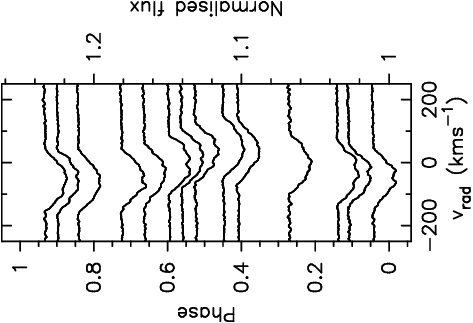}}}
\end{center}
\caption{\label{hd209295fase} 
Same as Fig.\,\ref{hd34025fase}, but for \hd209295.
The dotted line in the top left panel indicates the time of periastron passage.
}
\end{figure*}

\begin{figure*}[h]
\begin{center}
\resizebox{4.95cm}{!}{\rotatebox{270}{\includegraphics{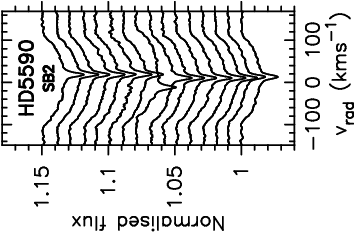}}}
\resizebox{4.15cm}{!}{\rotatebox{270}{\includegraphics{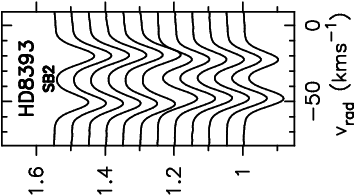}}}
\resizebox{4.15cm}{!}{\rotatebox{270}{\includegraphics{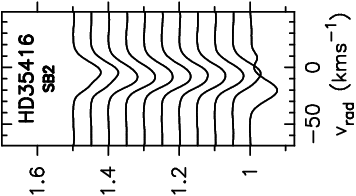}}}
\resizebox{4.95cm}{!}{\rotatebox{270}{\includegraphics{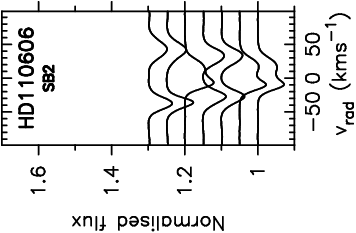}}}
\resizebox{4.15cm}{!}{\rotatebox{270}{\includegraphics{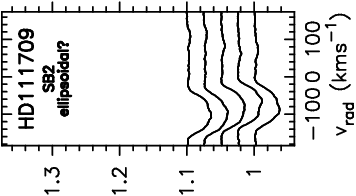}}}
\resizebox{4.15cm}{!}{\rotatebox{270}{\includegraphics{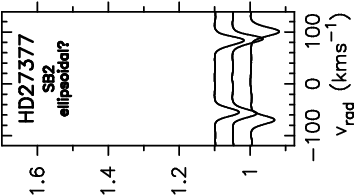}}}
\resizebox{4.95cm}{!}{\rotatebox{270}{\includegraphics{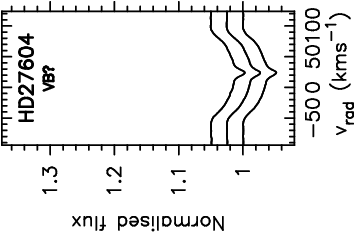}}}
\resizebox{4.15cm}{!}{\rotatebox{270}{\includegraphics{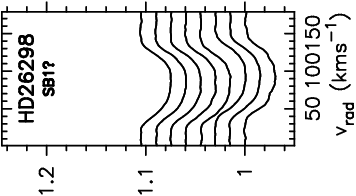}}}
\resizebox{4.15cm}{!}{\rotatebox{270}{\includegraphics{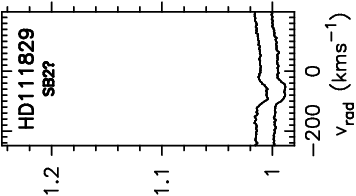}}}
\end{center}
\caption{\label{suspect} 
The cross-correlation profiles of the observed CORALIE spectra for the
(suspected) binary objects for which no orbital solution could be determined
with our current set of \coralie\ data.
Subsequent profiles are shifted in flux for clarity.
The name of the object and the orbital classification is given on top of each
panel.
``SB2'', ``SB1'', ``VB'' and ``ellipsoidal'' respectively indicate
double-lined binaries, single-lined binaries, visual binaries and ellipsoidal
variables. 
}
\end{figure*}

\setcounter{figure}{11}

\setcounter{figure}{12}

\begin{figure*}[h]
\begin{center}
\resizebox{7.42cm}{!}{\rotatebox{270}{\includegraphics{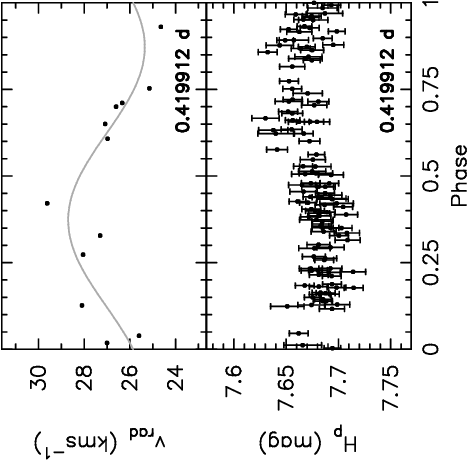}}}
\resizebox{5.20cm}{!}{\rotatebox{270}{\includegraphics{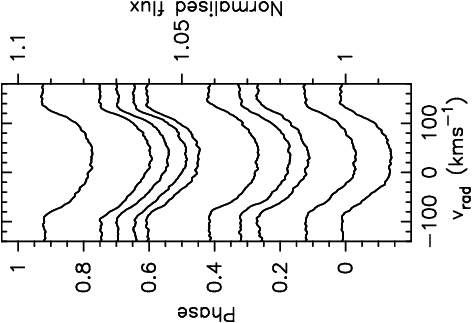}}}
\end{center}
\caption{\label{hd41448fase} 
Same as Fig.\,\ref{hd40745fase}, but for \hd41448.
}
\end{figure*}

\begin{figure*}[h]
\begin{center}
\resizebox{7.42cm}{!}{\rotatebox{270}{\includegraphics{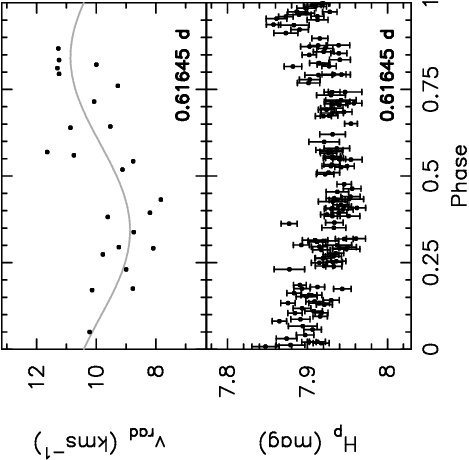}}}
\resizebox{5.20cm}{!}{\rotatebox{270}{\includegraphics{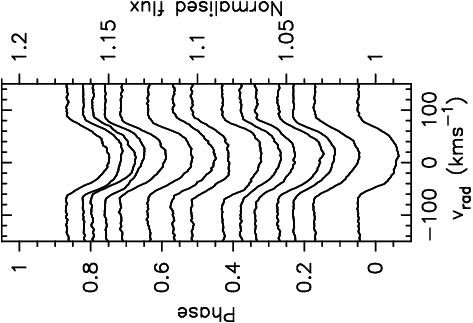}}}
\end{center}
\caption{\label{hd112685fase} 
Same as Fig.\,\ref{hd40745fase}, but for \hd112685.
}
\end{figure*}

\begin{figure*}[h]
\begin{center}
\resizebox{7.42cm}{!}{\rotatebox{270}{\includegraphics{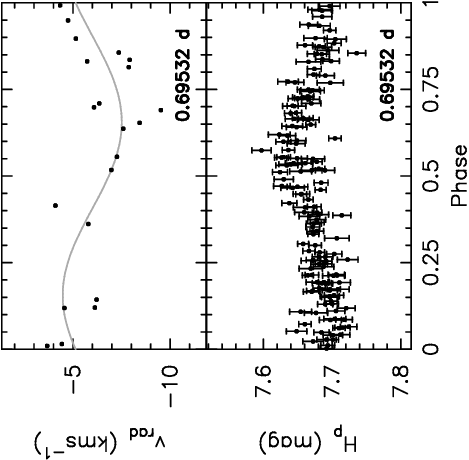}}}
\resizebox{5.20cm}{!}{\rotatebox{270}{\includegraphics{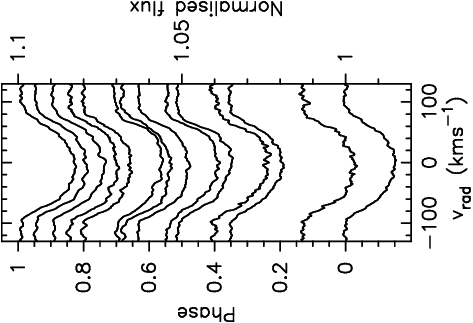}}}
\end{center}
\caption{\label{hd187028fase} 
Same as Fig.\,\ref{hd40745fase}, but for \hd187028.
}
\end{figure*}

\begin{figure*}[h]
\begin{center}
\resizebox{7.42cm}{!}{\rotatebox{270}{\includegraphics{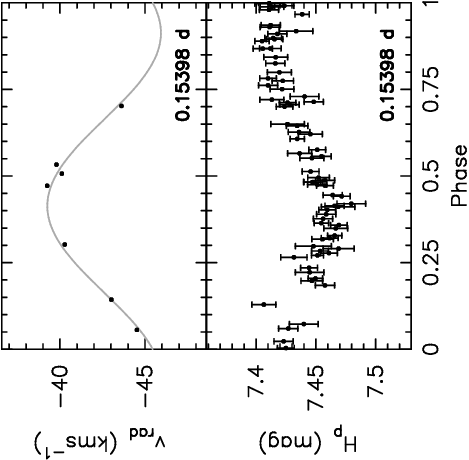}}}
\resizebox{5.20cm}{!}{\rotatebox{270}{\includegraphics{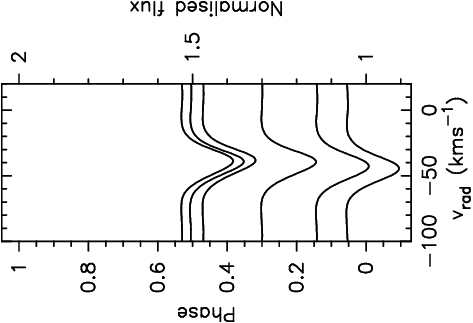}}}
\end{center}
\caption{\label{hd125081fase} 
Same as Fig.\,\ref{hd40745fase}, but for \hd125081.
}
\end{figure*}

\begin{figure*}[h]
\begin{center}
\resizebox{7.42cm}{!}{\rotatebox{270}{\includegraphics{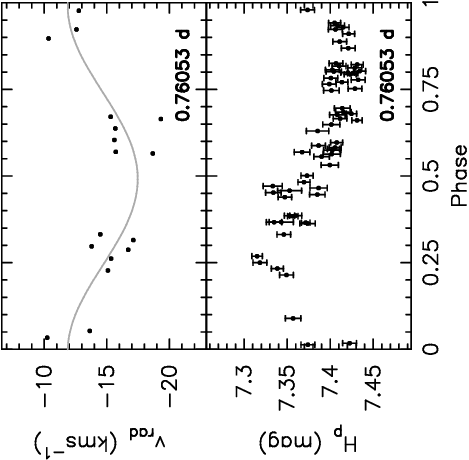}}}
\resizebox{5.20cm}{!}{\rotatebox{270}{\includegraphics{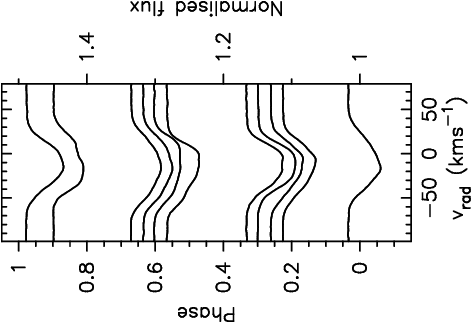}}}
\end{center}
\caption{\label{hd135825fase} 
Same as Fig.\,\ref{hd40745fase}, but for \hd135825.
}
\end{figure*}

\begin{figure*}[h]
\begin{center}
\resizebox{7.42cm}{!}{\rotatebox{270}{\includegraphics{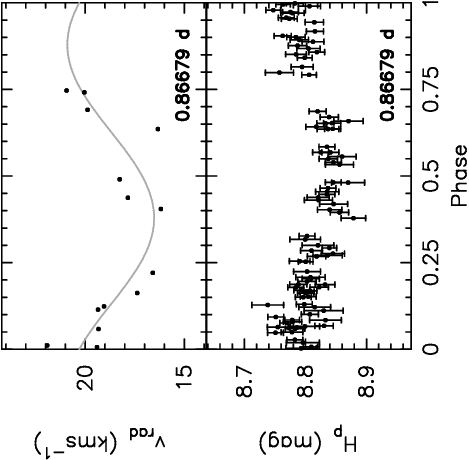}}}
\resizebox{5.20cm}{!}{\rotatebox{270}{\includegraphics{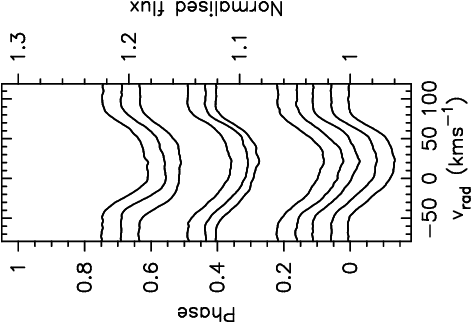}}}
\end{center}
\caption{\label{hd218225fase} 
Same as Fig.\,\ref{hd40745fase}, but for \hd218225.
}
\end{figure*}

\setcounter{figure}{19}

\setcounter{figure}{20}

\begin{figure*}
\begin{center}
\resizebox{7.42cm}{!}{\rotatebox{270}{\includegraphics{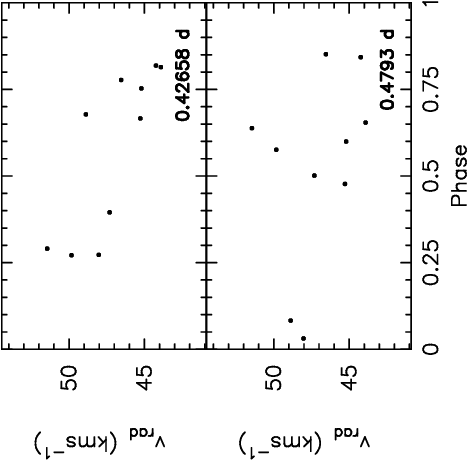}}}
\resizebox{5.20cm}{!}{\rotatebox{270}{\includegraphics{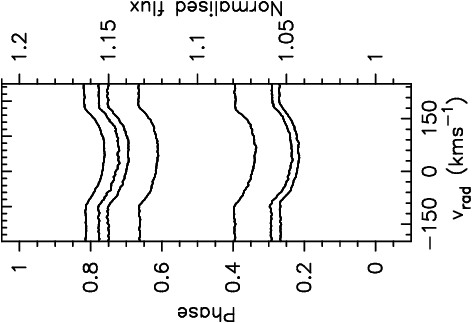}}}
\end{center}
\caption{\label{hd149989fase} 
Phase diagrams of the radial velocity \vrad\ with two known \hipparcos\
periods of \hd149989 (left).
The period values are given in the bottom right corner.
The reference epoch is HJD 2450000.
In the right panel, a selection of observed cross-correlation profiles are
shown as a function of the main \hipparcos\ period. 
}
\end{figure*}

\begin{figure*}
\begin{center}
\resizebox{7.42cm}{!}{\rotatebox{270}{\includegraphics{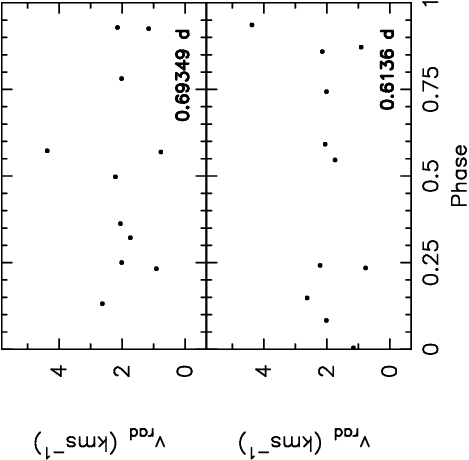}}}
\resizebox{5.20cm}{!}{\rotatebox{270}{\includegraphics{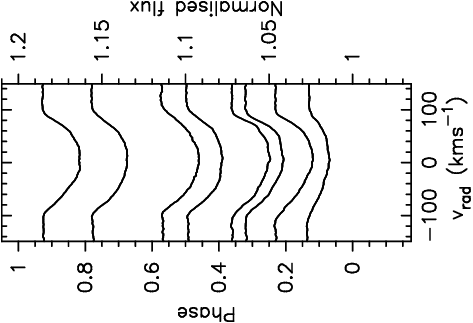}}}
\end{center}
\caption{\label{hd216910fase}
Same as Fig.\,\ref{hd216910fase}, but for \hd216910. 
}
\end{figure*}

\begin{figure}[h]
\begin{center}
\resizebox{4.5cm}{!}{\rotatebox{270}{\includegraphics{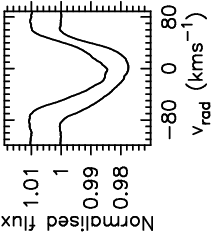}}}
\end{center}
\caption{\label{CPVspreidingplot} 
The cross-correlation profiles of the observed CORALIE spectra for \hd65526.
Subsequent profiles are shifted in flux for clarity.
}
\end{figure}

\begin{figure*}[h]
\begin{center}
\resizebox{4.95cm}{!}{\rotatebox{270}{\includegraphics{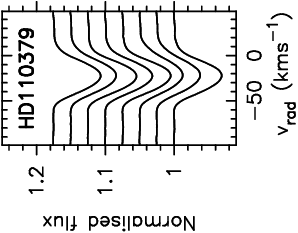}}}
\resizebox{4.15cm}{!}{\rotatebox{270}{\includegraphics{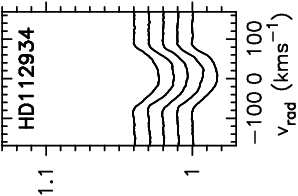}}}\\
\resizebox{4.95cm}{!}{\rotatebox{270}{\includegraphics{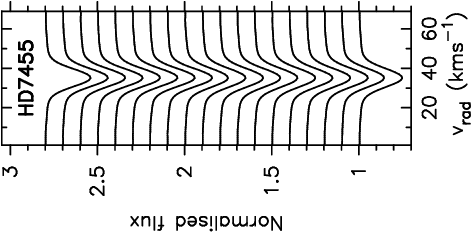}}}
\resizebox{4.15cm}{!}{\rotatebox{270}{\includegraphics{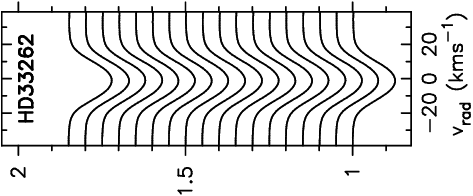}}}
\resizebox{4.15cm}{!}{\rotatebox{270}{\includegraphics{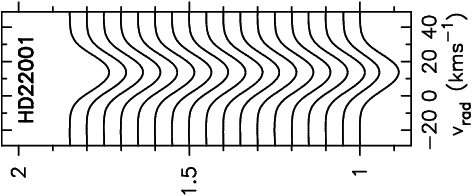}}}
\end{center}
\caption{\label{noCPVspreidingplot} 
The cross-correlation profiles of the observed CORALIE spectra for the objects
for which no clear correlation profile variations have been detected in our current
set of \coralie\ data. 
Subsequent profiles are shifted in flux for clarity.
The HD number of the object is given on top of each panel.
}
\end{figure*}

\end{document}